\pdfoutput=1
\documentclass[
  twocolumn,
  prd,
  showpacs,
  amsmath,
  amssymb,
  superscriptaddress
]{revtex4-1}

\usepackage{bm}
\usepackage{enumerate}
\usepackage{graphics}
\usepackage{hyperref}
\usepackage{amssymb}
\usepackage{color}
\usepackage{braket}
\usepackage{pifont}
\usepackage{empheq}
\hypersetup{colorlinks=true}
\usepackage{scalerel,stackengine}
\stackMath
\newcommand\reallywidehat[1]{%
\savestack{\tmpbox}{\stretchto{%
  \scaleto{%
    \scalereL_*[\widthof{\ensuremath{#1}}]{\kern-.6pt\bigwedge\kern-.6pt}%
    {\rule[-\textheight/2]{1ex}{\textheight}}
  }{\textheight}%
}{0.5ex}}%
\stackon[1pt]{#1}{\tmpbox}%
}

\newcommand{\dg}{^\dagger}

\newcommand{\pmat}[1]{\begin{pmatrix}#1\end{pmatrix}}

\begin{document}

\title{Renormalization group analysis for the quasi-1D superconductor BaFe$_2$S$_3$}

\author{Elio J.\ K\"onig}
\affiliation{Department of Physics and Astronomy, Rutgers University, Piscataway, New Jersey, 08854, USA}

\author{Alexei M.\ Tsvelik}
\affiliation{Condensed Matter Physics and Materials Science Division, Brookhaven National Laboratory, Upton, NY 11973-5000, USA}

\author{Piers Coleman}
\affiliation{Department of Physics and Astronomy, Rutgers University, Piscataway, New Jersey, 08854, USA}
\affiliation{Department of Physics, Royal Holloway, University of London, Egham, Surrey TW20 0EX, UK}
\date{\today}

\begin{abstract}
Motivated by 
the discovery of superconductivity in the two-leg, quasi-one
dimensional  ladder compound, 
BaFe$_2$S$_3$
we present a 
renormalization group study of electrons moving on 
a two leg,
two orbital ladder,  subjected to Hubbard repulsion $U$ and
Hund's coupling $J$.  
In our calculations, 
we adopt tightbinding parameters obtained from ab-initio studies on this
material. At incommensurate filling, the long wavelength analysis
displays four phases as a function of $0 \leq J/U <1$. 
We show that a 
fully gapped
superconductor is stabilized at sufficiently large Hund's coupling,
the relative phases at the three Fermi points are ``+,-,-''. By
contrast, when the system is tuned to half filling, Umklapp scattering
gives rise to Mott insulating phases.  We discuss the general 
implications of our study for the broad class of iron-based
superconductors. 
\end{abstract}


\maketitle

\section{Introduction}

The origin of the superconducting phase of iron pnictides and iron
chalcogenides remains an open and fascinating  puzzle. The 
robust nature of iron-based superconductivity,
found in both 
tetrahedral iron-pnictide and iron-selenide structures, despite a wide
variation in Fermi surface geometries and crystal structures, is
particularly striking\cite{Paglione:2010ct,Hirschfeld:2011jy,Chubukov_review2012,Yi:2017dg}. 
The transition temperatures appear to be broadly independent of whether 
the particular {compound displays}
hole pockets, electron pockets or both types of carrier. {In view of substantial on-site Coulomb repulsion, this makes the quest for a generic pairing mechanism particularly challenging.}
Moreover, superconductivity has been reported in 
systems in both tetragonal and orthorhombic phases.

\begin{figure}
\includegraphics[scale=.21]{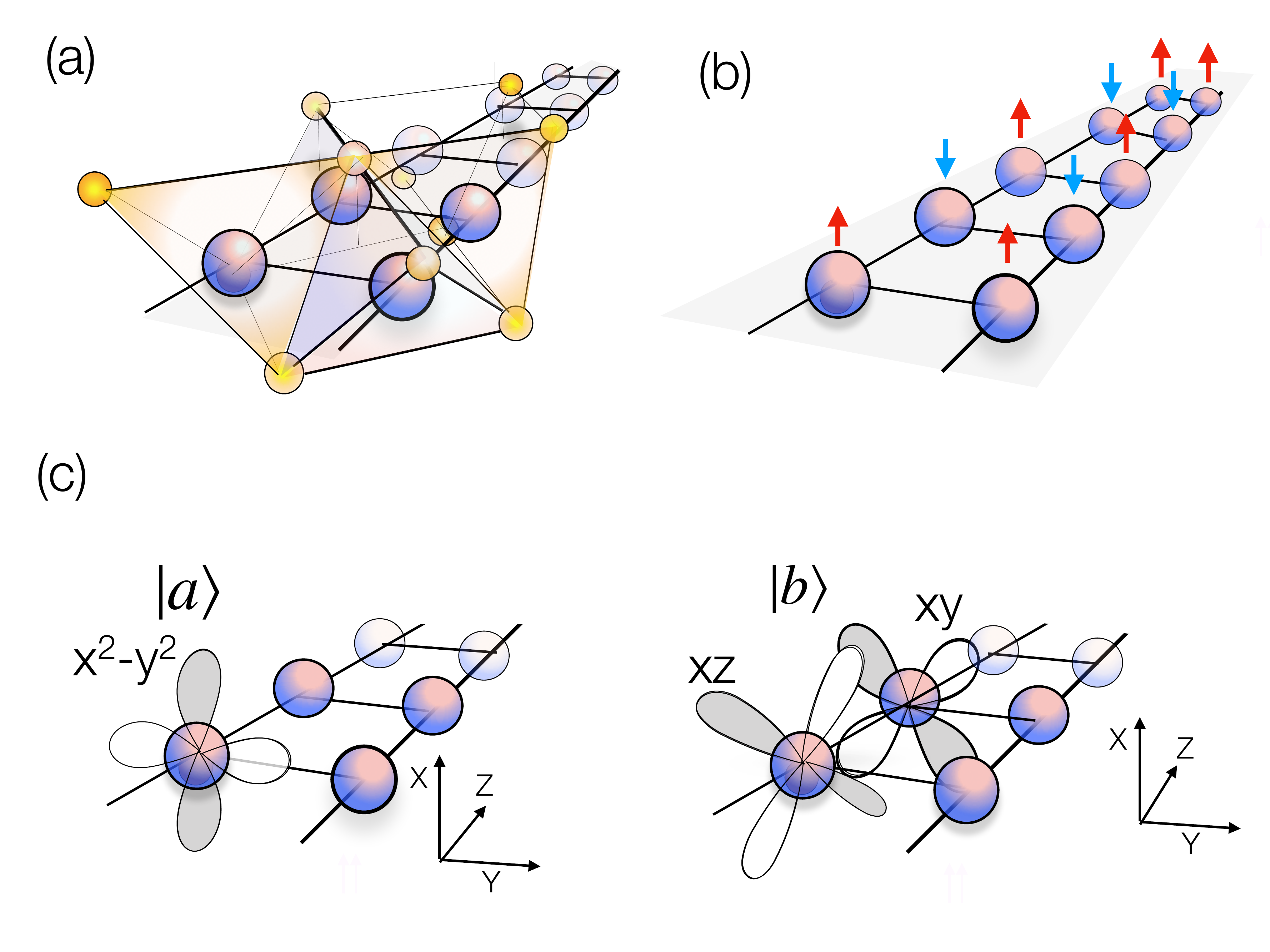}
\caption{ (a) Ladder structure of BaFe$_2$S$_3$ shown running along $Z$ axis,
with sulphur tetradhedra (gold)
surrounding iron atoms (blue).
(b) Staggered antiferromagnetic ``stripe'' structure along ladder
in magnetic phase at ambient pressure (c)
Following ab-initio
calculations~\cite{AritaSuzuki2015, PatelDagotto2016} we consider an
effective two-orbital tight model composed from the $d_{x^2-y^2}\equiv
\vert  a\rangle $ orbital 
and a $\vert b\rangle $ orbital composed of a 
superposition of $d_{xz}$ and $d_{xy}$ orbitals. 
}
\label{fig:LadderOrbitals}
\end{figure}

A particularly exotic example of this superconducting 
robustness within the zoo of iron-based materials is BaFe$_2$S$_3$ under
pressure~\cite{TakahashiOhgushi2015}.
While the sulfide shares the same staggered tetrahedral structure
as its quasi-two dimensional (q2D) cousins, with 
a band of delocalized d-electrons forming between Fe$^{2+}$ ions,  
here the tetrahedra are organized into two-leg ladders, 
forming a quasi-one dimensional (q1D) structure
(see Fig.~\ref{fig:LadderOrbitals}).   {One of the
fascinating 
aspects of this system, is that it opens up the possibility of
analyzing the physics of iron based superconductivity using the
powerful tools of one-dimensional renormalization group and
bosonization. }

The experimentally observed phase
diagram~\cite{YamauchiOhgushi2015,ChiOhgushi2016,MaterneAlp2018} resembles that of q2D
materials, with a superconducting dome developing at the end point of an
antiferromagnetic (AF) phase. A further analogy is the "stripe"
ordering of spins, in the ladder system this corresponds to a
ferromagnetic ordering on the rungs and AF ordering along the
legs~\cite{TakahashiOhgushi2015}. On the other hand, a major difference
between BaFe$_2$S$_3$ and the q2D materials is the insulating nature
of the antiferromagnet, which contrasts with the bad metal behavior in
more conventional materials.
Three years after the discovery of superconductivity in
BaFe$_2$S$_3$ theoretical studies of this novel superconductor 
are still relatively sparse. Two
groups~\cite{AritaSuzuki2015, PatelDagotto2016} reported ab-initio
calculations extracting an effective two orbital tight
binding model. The dispersion relation in ladder direction and the
orbital content near the Fermi surface (see also
Fig.~\ref{fig:DispRel})~\footnote{The plot of the dispersion relation
in Fig.~\ref{fig:DispRel} using the parameters from
Ref.~\cite{PatelDagotto2016} is inconsistent with Fig.~2 b) of the
same reference. In contrast to the original reference, we find that
the level crossing at $k \sim \pi/3$ is avoided. Note that Figs. 2a),
3 a) and 3 b) of Ref.~\cite{PatelDagotto2016} do report an avoided
level crossing near $\pi/3$ and are thus qualitatively equivalent to
our plot.} qualitatively agree in both studies. A rough summary of
energy scales follows: the Hubbard $U \sim 3 {eV}$, the bandwidth (intraladder
hopping) $\Lambda \sim 2 eV$, the Hund's coupling $J \sim 0.5 eV$,
the interladder hopping $ t_\perp \sim 0.25 eV$.  The effects of
interactions on this compound are
discussed in Ref.~\cite{AritaSuzuki2015} based on the analysis of the
Lindhard function, while the authors of Ref.~\cite{PatelDagotto2016}
investigate the interplay of Hubbard and Hund's coupling using density
matrix renormalization group. Both papers correctly reproduce the
"striped" AF state, which was also reported in earlier density
functional theory studies~\cite{SuzukiIkeda2015}. A slave spin approach~\cite{PizarroBascones2018} based on the tight-binding model of Ref.~\cite{AritaSuzuki2015} reveals orbital selective correlations. The first order,
pressure induced magnetic transition was recently
scrutinized~\cite{ZhangDong2017} for BaFe$_2$S$_3$ and related
materials. We note that stripe order may also be understood in terms
of a $J_1$ - $J_2$ AF Heisenberg model: As soon as the diagonal
coupling $J_2$ exceeds $J_1/2$,  stripe order is energetically
favored over a N\'eel-type order\cite{ccl}. Superconductivity is discussed
qualitatively in Ref.~\cite{AritaSuzuki2015} and the calculation of
Ref.~\cite{PatelDagotto2016} indicates a pairing tendency in hole
doped systems at sufficiently strong interactions. In a recent
follow-up~\cite{PatelDagotto2017}, the same group reports pairing
tendencies in single-leg chains with the same orbital content and the
importance of Hund's coupling was emphasized. It is important to
realize, that density matrix renormalization group studies for
multi-orbital Hamiltonian are numerically costly and thus restricted
to small systems.

\begin{figure}
\includegraphics[scale=.52]{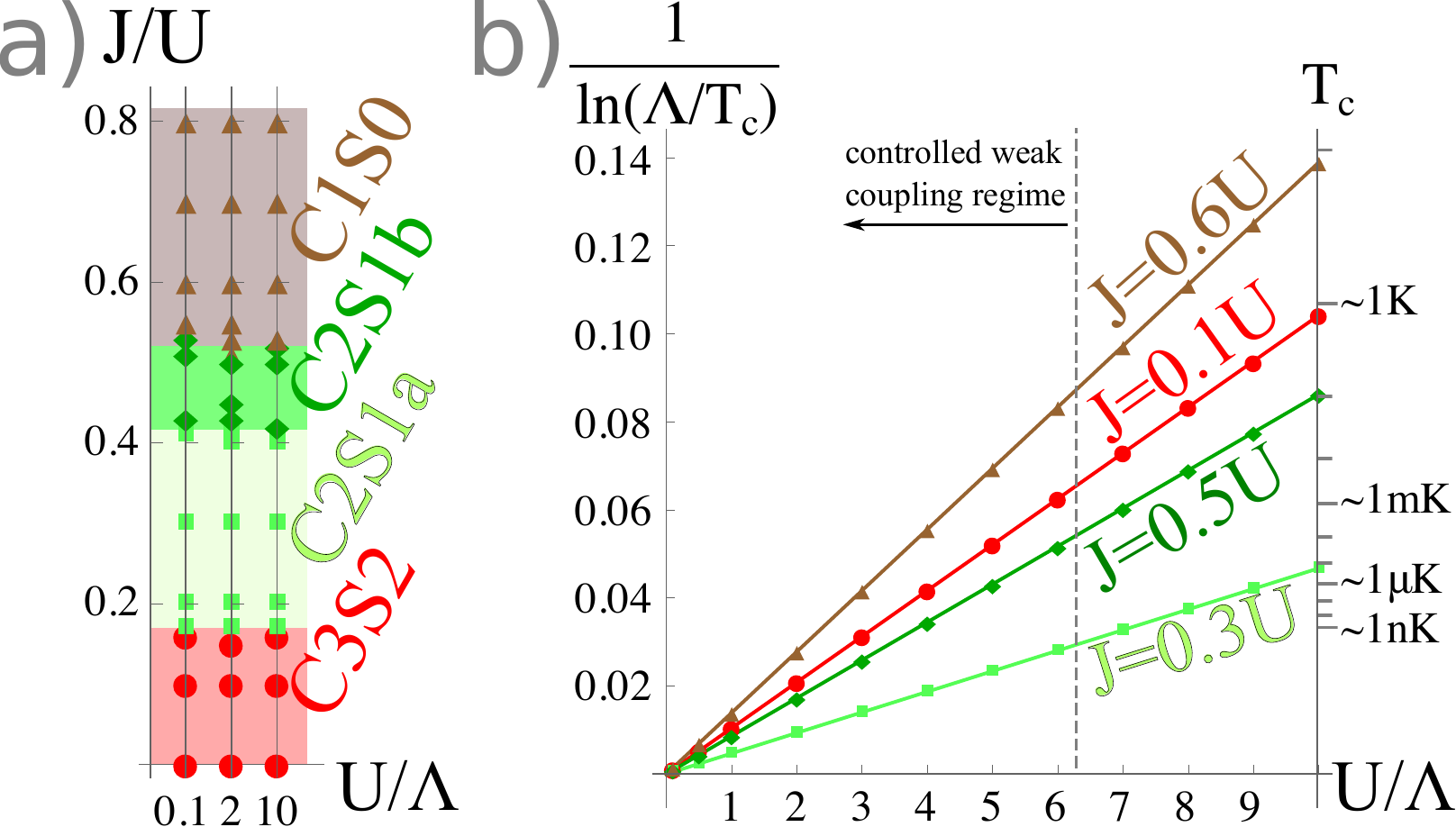} 

\caption{Phase diagram and critical temperature as a function of U. As
explained in the main text, the nature of the ground state depends on
the ratio $J/U$, only. The numerical solutions of the RG equations
presented here were obtained for Fermi velocities as defined by
Fig.~\ref{fig:DispRel}. Panel a): Phase diagram for three values of
$U/\Lambda$, data points are marked by dots (C3S2), squares (C2S1a),
diamonds (C2S1b) and triangles (C1S0) corresponding to four different
phases in ascending order of $J/U$. Panel b): The critical temperature
$T_c(U)$ associated to the instability of RG equations. Straight solid
lines are obtained using Eq.~\eqref{eq:Tc} with $T_c(\Lambda)$ as the
only numerical input, additional numerical solutions are again
presented as dot, square, diamond and triangle, respectively. We used
$\Lambda = \bar v/\tilde a$ and assumed $\Lambda \sim 1 eV$ to
estimate $T_c$ in Kelvin on the right vertical axis.}
\label{fig:PhaseDiagram}
\end{figure}
Motivated by these recent experimental and theoretical advances, here we
present a weak-coupling renormalization group study of  a
two-orbital, two-leg ladder with on-site Hubbard and Hund
interactions. The orbital content of Fermi-surface excitations is
chosen in accordance with
Refs.~\cite{AritaSuzuki2015,PatelDagotto2016} and we employ the same 
tight binding parameters as in~\cite{PatelDagotto2016} to
determine the Fermi velocities {at three pairs of Fermi momenta.}
{A crucial step in our approach, is the reformulation 
of the
onsite Coulomb and Hund's interactions as a set of 
eighteen interaction parameters
$g_{\mu}$ ($\mu\in (1,18)$) defining the  
strength of interaction 
between the left and right moving spin and charge currents in the various orbitals. } 
From this formulation, we are able to construct a set of coupled RG equations
that have the general form 
\begin{equation}
\frac{d g_\mu(y)}{d y} = \beta_{\mu \nu \rho}g_\nu(y)g_\rho(y), \label{eq:RGMaintext}
\end{equation}
{where $y =\ln {\Lambda}$ is the logarithm of the energy cut-off}. 
Our analysis of coupled RG equation{s}
for the ladder model is the 1D analog of the parquet-RG approach to
q2D iron based
superconductors~\cite{MaitiChubukov2010,ChubukovFernandes2016,ClassenChubukov2017}
and, being a weak coupling - long wavelength study it provides a
complimentary perspective to the strong coupling DMRG computations on finite
ladders~\cite{PatelDagotto2016}.

\begin{figure}
\includegraphics[scale=.52]{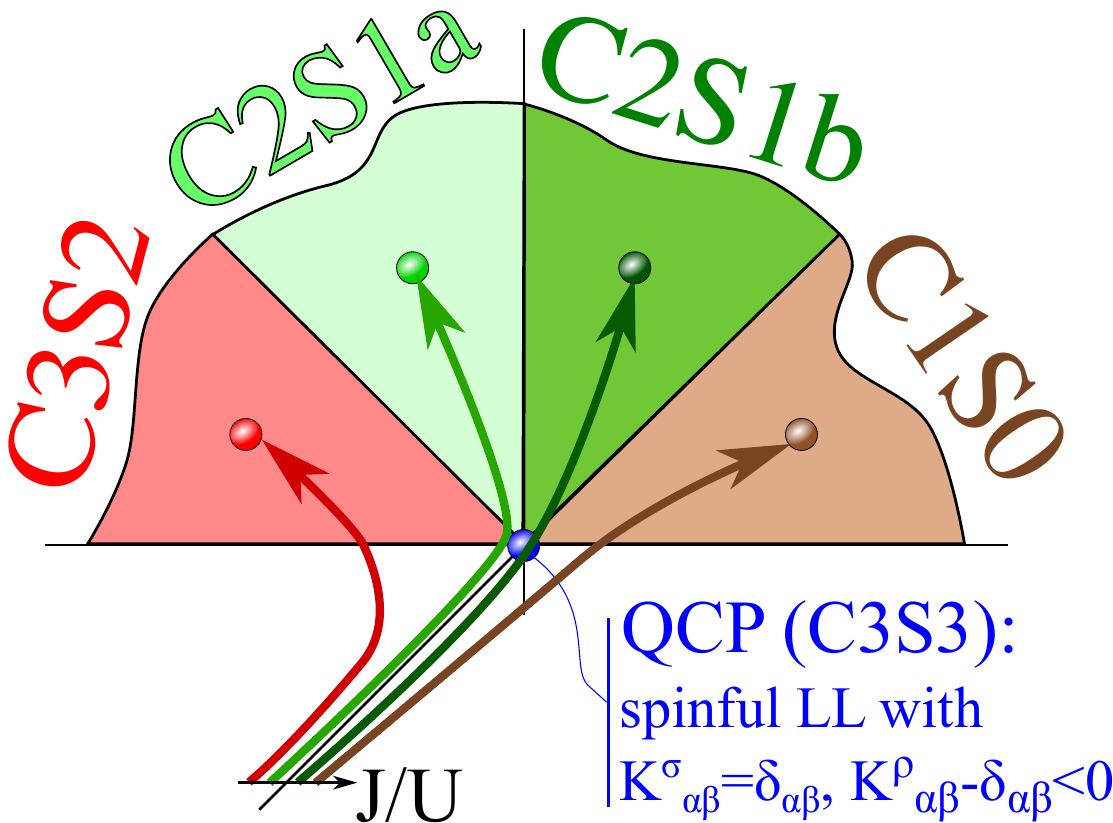} \caption{Schematic
summary of the RG flow obtained numerically in
Figs.~\ref{fig:U08J01_RG}-\ref{fig:U08J06_RG} of the Appendix. At the
initial stage, coupling constants flow towards a quantum critical
point (QCP), which in the present 1D study is just a spinful Luttinger
liquid (LL) with three charge and three spin modes (C3S3). Near the
QCP, the flows diverge towards four possible attractive fixed points
(phases), discussed in Fig.~\ref{fig:PhaseDiagram}. At intermediate
$J/U$, the RG flow approaches the QCP very closely, it therefore slows
down and $T_c$ is suppressed in the phases C2S1a and C2S1b. }
\label{fig:SchematicRG}
\end{figure}

{The key result of our paper is the identification of four stable ground-state phases which fan out from a central
quantum critical point described by a gapless Luttinger liquid {(LL)}. 
{Crucially, it is the strength of the 
Hund's coupling which tunes between the four phases. 
({see} figures ~\ref{fig:PhaseDiagram} and
~\ref{fig:SchematicRG}).
{Moreover, by analyzing the 
effect of {u}mklapp scattering at half-filling we can show that these
phases emerge from a Mott insulating phase that develops at half-filling. }

To characterize the excitation spectrum of these phases, we have
used bosonization to perform a semi-classical strong-coupling analysis
which permits a characterization of the gapless modes that dominate
the quasi-long-range order. 
Following the convention of 1D ladder systems,
we use the notation ``CmSn'' to describe a phase with m gapless
boson modes in the charge sector and n gapless modes in the spin
sector. The four stable phases can be summarized as follows: 
\begin{itemize}
\item C1S0 ($J/U > 0.55$) {Orbitally selective }{singlet
s}uperconductor,  {characterized by strong 
intraband pairing with relative phases $+,-,-$ on the three Fermi momenta.}
\item C2S1b ($J/U \in [0.45,0.55]$)  {Singlet, equal sign, intraband
superconductor at two Fermi momenta decoupled from a {LL}
at the third Fermi momentum.}
\item  C2S1a ($J/U\in [0.18,0.42]$)  Orbitally ordered charge density wave {decoupled from a LL}.
\item  C3S2 ($J/U\leq 0.18$) {Long range superconducting and charge density wave correlations stemming from one out of the three Fermi momenta.}
\end{itemize}

The structure of the paper is as follows: The microscopic model is
introduced in Sec.~\ref{sec:Model}, the RG analysis and the discussion
of the rich phase diagram are presented in Sec.~\ref{sec:RG}. We
conclude with a summary and outlook. Full details on the model and the RG
calculation are included in the appendices \ref{app:Micro} and
\ref{app:RG}.

\section{Model}
\label{sec:Model}

In this section we present the model under investigation. At each rung
of the ladder, see Fig.~\ref{fig:LadderOrbitals}, 
there are eight
degrees of freedom:  the chain index $\tau = 1,2$, the orbital
quantum number $\gamma = a,b$ and the spin z-component $\sigma =
\uparrow, \downarrow$. The corresponding electron annihilation operator at site $j$
is then written $d_{\tau \gamma\sigma } (j)$. We incorporate the chain and orbital indices
into a four-component spinor, defined for each site $j$ and spin
component $\sigma $ as follows:
\begin{equation}
 d_{\sigma} ({j}) =  
\pmat{
d_{1a\sigma }({j})  \\
d_{1b\sigma }({j}) 
\\ 
d_{2a\sigma }({j}) \\
d_{2b\sigma }({j}) 
}.
 \end{equation} 
\label{eq:DefAnnihilOp}
\subsection{Kinetic term} 
The dispersion relations presented in
Refs.~\cite{AritaSuzuki2015,PatelDagotto2016} are qualitatively
similar to Fig.~\ref{fig:DispRel}. For details of the tight binding
model see Appendix~\ref{app:tightbinding}.  
The important, robust
features of the model are as follows:
\begin{enumerate}[1)]

\item In the interval $k \in
[0,\pi]$, there are two Fermi points of right movers at $k_{\rm I}, k_{\rm II}$
and one left mover at {$- k_{\rm III}$}. For each of
them, time reversal symmetry imposes Fermi points of opposite velocity
at the reversed momentum. 

\item 
The excitations near $k_{\rm II}$ and
$ k_{\rm III}$ are even parity under reflections in the mirror
plane running along the ladder ($Y\rightarrow -Y$), whereas the
excitations near $k_{\rm I}$ are odd-parity under this reflection.

\item {The} excitations near $ \pm k_{\rm II}$ or $\pm
k_{\rm III}$ {approximate} pure orbital states ${a}$ and ${b}$ respectively, 
while excitations
near $ \pm k_{\rm I}$ are in an orbital superposition $(\ket{a} \pm i
\ket{b})/\sqrt{2}$. 

\item At half-filling the integration over filled states implies $k_{\text{I}}+k_{\text{II}}+k_{\text{III}}=
0$ (we took into account that crystal momenta of $0$ and $2\pi$ are equivalent).

\end{enumerate}

In the continuum limit, this motivates an expansion in low-energy
modes. For convenience, we label the continuous position along the
chain by $x_{j} = \tilde a j$ and set the
lattice constant $\tilde{a} = 1$ everywhere in the paper. (Note that
by using $x$, we have tacitly rotated the co-ordinate system relative
to Fig.~\ref{fig:LadderOrbitals} and introduced $(x,y,z) = (Z,X,Y)$). 
In the continuum limit, we can factor out the rapidly varying
components of the electron field, and decompose it into right ($R$)
and left ($L$) moving components as follows:
\begin{equation}\label{eq:LowEnExp}
\hat d_{\sigma} (x)= \sum_{{\beta }\in [{\rm I,III}]}\left[
e^{ik_{{\beta }}x}\Psi_{{\beta }}\hat a^{R}_{{\beta}\sigma }(x)
+ 
e^{-ik_{{\beta }}x}\Psi^{*}_{{\beta }}\hat a^{L}_{{\beta }\sigma }(x)
\right]
\end{equation}
where $a_{{\beta }\sigma}^{R,L}(x)$ corresponding to right and left-moving
components of the fields and the three Fermi momenta are 
$k_{{\beta }}= (k_{{\rm I}}, k_{\rm {II}}, k_{\rm {III}})$. 
The spinor components of the wavefunctions are

\begin{eqnarray}\label{l}
\Psi_{{\rm I}} &=&
 \left (\begin{array}{c} 1/\sqrt{2} \\ -1/\sqrt{2} \end{array} \right
)_\tau \otimes \left (\begin{array}{c} 1/\sqrt{2} \\
i/\sqrt{2} \end{array} \right )_\gamma 
= \pmat{\ 1/2\\
\ i/2 \\
-1/2\\
-i/2}, \label{eq:PsiI}\cr
\Psi_{{\rm II}} &=&   \left (\begin{array}{c} 1/\sqrt{2} \\ 1/\sqrt{2} \end{array} \right
 )_\tau \otimes \left (\begin{array}{c} 1\\0 \end{array} \right
 )_\gamma 
 = \pmat{{1}/{\sqrt{2}}\\ 0\\{1}/{\sqrt{2}}\\ 0},\cr
\Psi_{\rm III} &=& \left (\begin{array}{c} 1/\sqrt{2} \\
1/\sqrt{2} \end{array} \right )_\tau \otimes \left (\begin{array}{c}
0\\1 \end{array} \right )_\gamma 
= \pmat{0\\
{1}/{\sqrt{2}}\\ 0\\{1}/{\sqrt{2}}}.
\end{eqnarray}
The kinetic part of the long-wavelength Hamiltonian is then
\begin{equation}\label{l}
H_{{\rm kin}} = -iv_{{\beta }}
\int dx 
\left[{a^{R}_{{\beta}\sigma }}\dg \nabla_{x}a^{R}_{{\beta}\sigma }-  {a^{L}_{{\beta}\sigma
}}\dg \nabla_{x}
a^{L}_{{\beta}\sigma }\right].
\end{equation}
where we use an index notation for summation over the repeated
variables $\beta =  {\text{(I, II, III)}}$ and
$\sigma = (\uparrow,\downarrow )$.
The Fermi-velocities $v_{\rm I, II, III}$ as well as the values
$k_{\rm I, II, III}$ are non-universal and may continuously vary as a
function of the  microscopic parameters. 
\begin{figure}
\includegraphics[scale=.5]{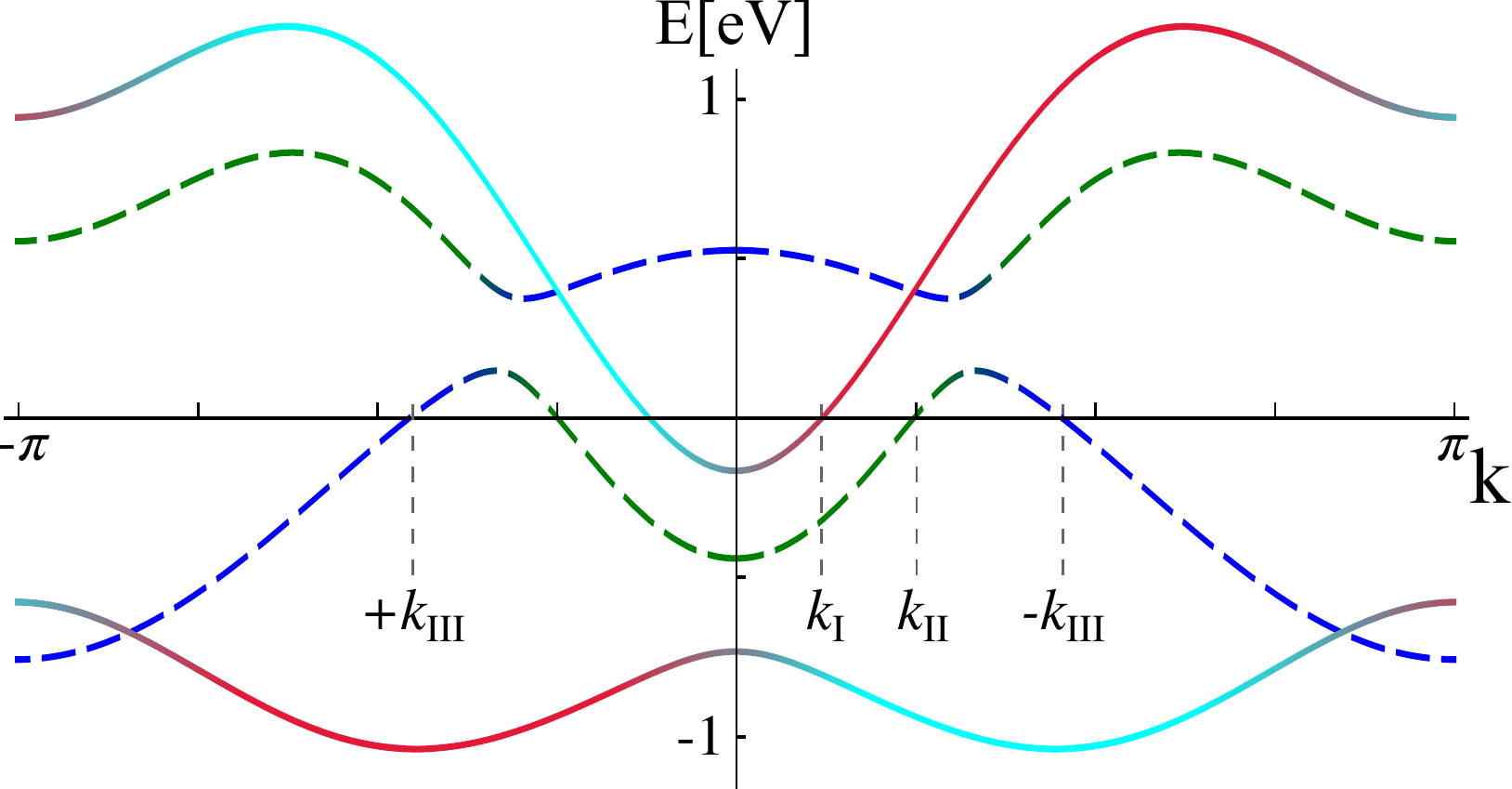} 
\caption{Dispersion relation using the tight binding parameters determined in Ref.~\cite{PatelDagotto2016} for an applied pressure of 12.36 GPa and including next-nearest neighbor hopping. The dashed (solid) curves correspond to states which are symmetric (antisymmetric) under $y \rightarrow - y$ reflection and at the avoided crossing at $k \approx \pi/3$ ($k = 0$) the orbital character of the eigenstates changes. For the symmetric states, we use dark green to represent orbital $\ket{a}$ and blue for $\ket{b}$. For the antisymmetric states the superposition $[\ket{a} + i \ket{b}]/\sqrt{2}$ ($[\ket{a} - i \ket{b}]/\sqrt{2}$) is represented in red (cyan).}
\label{fig:DispRel}
\end{figure}

\subsection{Interaction terms}

We assume a simplified model of onsite Hubbard and Hund interactions,
$H_{\rm int}= \sum_{{j}}H_{U} ({j}) + H_{J} ({j})$, where the interactions
at each site are
\begin{subequations}
\label{eq:IAtightbinding}
\begin{align}
H_{U} ({j})&= \frac{U}{2}  { \sum_{\substack{\tau,  {\gamma, \sigma,}\\ { \gamma', \sigma'}}}}' n_{ \tau \gamma\sigma} ({j}) n_{\tau \gamma' \sigma'} ({j}) , \label{eq:Hubbard}\\
H_{J} ({j}) &= - 4 J \sum_{\tau } \vec S_{ \tau a} ({j}) \cdot \vec S_{ \tau b} ({j}) ,
\end{align}
\end{subequations}
\noindent 
where {the summation symbol $\sum '$ excludes $(\gamma, \sigma) = (\gamma', \sigma')$ and }we have introduced density and spin operators, defined as follows:
\begin{eqnarray}\label{l}
n_{\tau \gamma \sigma} ({j})&=&
d_{\tau\gamma \sigma}^\dagger {(j)} d_{\tau \gamma \sigma} ({j}),\\
\vec S_{\tau \gamma} ({j}) &=& \sum_{\sigma,\sigma'}d_{
\tau \gamma \sigma}\dg ({j}) 
\left( \frac{\vec \sigma}{2}\right)
_{\sigma, \sigma'}
d_{\tau \gamma \sigma'} ({j}).
\end{eqnarray}
More complicated Hubbard-Kanamori interactions with non-equal intra- and interorbital repulsion do not alter the main conclusions of our study and are therefore discussed in Apps.~\ref{app:Interaction} and \ref{app:NonEqualHubbard}. In the continuum limit, it is useful
to represent the interaction term in terms of scalar 
\begin{equation}\label{l}
{\cal J}^r _{\alpha \beta } (x) = \sum_\sigma a_{\alpha  \sigma}^{r} (x)\dg a_{\beta 
\sigma}^{r} (x), \qquad (r\in \{R,L \})
\end{equation}
and vector currents 
\begin{equation}\label{l}
\vec{\cal J}^r_{\alpha\beta} (x) =
\sum_{\sigma, \sigma'} {a_{{\alpha}
\sigma}^{r}}\dg (x)\left(
\frac{\boldsymbol{\sigma}}{2} \right)
_{\sigma, \sigma'}
a_{{\beta},
\sigma'}^{r} (x),
\end{equation}
involving states near the Fermi points ${\alpha, \beta} = \text{I,
II, III}$. Writing $H_{int }=\int dx {\cal H}_{int} (x)$ in terms of
the Hamiltonian density, then 
\begin{subequations}
\begin{align}
\mathcal H_{\rm int} (x) &= \tilde c_{\alpha\beta}^\rho {\cal J}^R _{\alpha\beta} (x) {\cal J}^L _{\alpha\beta} (x) + \tilde f_{\alpha\beta}^\rho {\cal J}^R_{{\alpha \alpha}} (x){\cal J}^L_{{\beta \beta}} (x) \notag \\
& -\left [\tilde c_{\alpha\beta}^\sigma \vec{\cal J}_{\alpha\beta}^R (x)  \cdot \vec{\cal J}_{\alpha\beta}^L (x)  + \tilde f_{\alpha\beta}^\sigma \vec{\cal J}_{{\alpha \alpha}}^R (x) \cdot \vec{\cal J}_{{\beta \beta}}^L (x) \right ],
\end{align}
where we have used an index summation on the indices ${\alpha, \beta}$.
The bare interaction constants 
are given by 
\begin{eqnarray}
 \tilde f_{\alpha\beta}^\rho &=& \frac{U}{8}\left (\begin{array}{ccc}
 0 & 3 & 3 \\ 
 3 & 0 & 4 \\ 
 3 & 4 & 0
 \end{array} \right )_{\alpha\beta}, \\
 \tilde f_{\alpha\beta}^\sigma &=& \frac{U}{2}\left (\begin{array}{ccc}
 0 & 1 & 1 \\ 
 1 & 0 & 0 \\ 
 1 & 0 & 0
 \end{array} \right ) _{\alpha\beta}+ J\left (\begin{array}{ccc}
 0 & 1 & 1 \\ 
 1 & 0 & 2 \\ 
 1 & 2 & 0
 \end{array} \right ) _{\alpha\beta} ,\\
\tilde c_{\alpha\beta}^\rho &=& \frac{U}{8}\left (\begin{array}{ccc}
 8 & 1 & 1 \\ 
 1 & 4 & 0 \\ 
 1 & 0 & 4
 \end{array} \right ) _{\alpha\beta} -\frac{J}{4}\left (\begin{array}{ccc}
 3 & 0 & 0 \\ 
 0 & 0 & 0 \\ 
 0 & 0 & 0
 \end{array} \right ) _{\alpha\beta}, \\
 \tilde c_{\alpha\beta}^\sigma &=& \frac{U}{2}\left (\begin{array}{ccc}
 0 & 1 & 1 \\ 
 1 & 4 & 0 \\ 
 1 & 0 & 4
 \end{array} \right ) _{\alpha\beta} + J\left (\begin{array}{ccc}
 1 & 0 & 0 \\ 
 0 & 0 & 0 \\ 
 0 & 0 & 0
 \end{array} \right ) _{\alpha\beta}{,}
\end{eqnarray}
\label{eq:bareinteraction}
\end{subequations}
see Appendix~\ref{app:Interaction} for derivation.

By convention~\cite{LinFisher1997}, the diagonal elements of forward
scattering amplitudes $\tilde f_{\alpha\beta}$ are chosen to vanish and are 
absorbed into the Cooper channel constants $\tilde c_{\alpha\beta}$. Completely
chiral interactions of the form $({\cal J}^R)_{\alpha \alpha } ({\cal J}^R)_{\beta \beta }$ 
will also be  disregarded since they are
not renormalized and do not 
renormalize the above couplings at one loop order. Formally, the itinerant approach assumes $\bar v = \sum_{\alpha} v_{\alpha}/3 \gg U/(2\pi)$ and we consider $U > J$.

\subsection{Umklapp scattering}
\label{sec:umklappMainText}

The effective low-energy Hamiltonian presented above is restricted
to two-body interactions, which are marginal operators. Interactions
involving a higher number of fermionic operators are generated during RG but are irrelevant near
the non-interacting fixed point and therefore usually disregarded. However, as the strong coupling
regime is approached such terms can become relevant. 

An example of such a three-body interaction that encodes qualitatively new physics is given by
umklapp scattering. It develops at commensurate filling, only, by means of the processes represented in Fig.~\ref{fig:Umklapp}. In the
present model umklapp scattering is a three body interaction even at half filling in
view of the aforementioned constraint $\sum_{\beta  = \text{I}}^{\text{III}} k_\beta  =
{ 0}$: Three right movers, one for each Fermi point, have to conspire and collectively transfer twice their total momentum (a crystal momentum) to the lattice. The difference between our model and more conventional Hubbard
models, where umklapp scattering is typically a two body interaction,
derives from the detailed band structure of Fig.~\ref{fig:DispRel}.
\begin{figure}
\includegraphics[scale=.75]{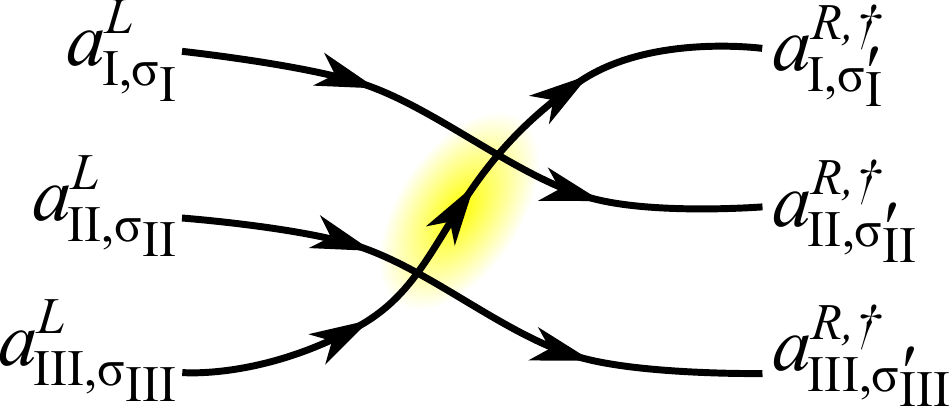} \caption{For the model
defined by Fig.~\ref{fig:DispRel}, the umklapp process at half filling
is a three body interaction. On the tree level, such processes are
generated by means of two-body interactions and involve one
virtual state away from the Fermi energy. Note that momentum is
conserved modulo an inverse lattice vector only. This diagram
indicates that the bare coupling is of the order $U^2/\bar v$.}
\label{fig:Umklapp}
\end{figure}

Umklapp scattering implies the following terms, see Fig.~\ref{fig:Umklapp}
\begin{equation}\label{eq:UmklappMainText}
\mathcal H_{\rm umklapp} \sim  - g_{U}
\left[
\prod_{\rm \beta = I}^{\rm III}
\left (
{a_{\beta  \sigma_\beta '}^{R\dagger }}
a_{\beta \sigma_{\beta}}^{L}\right ) 
e^{-i 2 k_\beta x}
+ \text{H.c.} \right]. 
\end{equation}
Note that momentum is conserved only modulo the reciprocal lattice vector $2\pi$.
For the present model, there is a multitude of such interactions which
differ by dissimilar spin indices (only the total spin is
conserved). All of them have bare coupling constants $g_U \sim U^2/\bar v$.

\section{RG analysis at incommensurate filling}
\label{sec:RG}

In this section we present the RG analysis for the model defined in the previous section. We first concentrate on the case of incommensurate filling.

The low-energy theory introduced above is analogous to that of a three
leg ladder~\cite{Arrigoni1996,KimuraAoki1996,Schulz1996}, albeit with
rather anisotropic interactions. It is known that the 18 independent
parameters $\tilde c_{\alpha\beta}^{\sigma,\rho}$ and $\tilde
f_{\alpha\beta}^{\sigma,\rho}$ form a closed set of running coupling constants
under one-loop RG. The RG equations for a generic N-leg ladder were
derived in Ref.~\cite{LinFisher1997}. Collecting the coupling
constants into a single eighteen component vector $g_\mu = (\tilde
c_{\alpha\beta}^{\rho},\tilde f_{\alpha\beta}^{\rho},\tilde c_{\alpha\beta}^{\sigma}, \tilde
f_{\alpha\beta}^{\sigma})_\mu$, the RG equations have the form {of Eq.~\eqref{eq:RGMaintext}}.
{Physically, the logarithmic scale factor $y = \ln(\Lambda/T)$ is
determined by } the ratio of the UV cut-off $\Lambda \sim \bar v$ and
temperature $T$. The detailed form of the structure factors
$\beta_{\mu \nu \rho}$ is given in App.~\ref{app:RG}. Typically, 
the coupling constants diverge at a characteristic scale
$T_c$. As a working definition we associate this scale with
the onset of a symmetry broken phase, bearing in mind that strictly speaking,
the divergence of one-loop RG in 1D marks the onset of the strong
coupling regime and the development of gaps in some parts of the spectrum. In
view of the simple structure of one-loop
equations~\eqref{eq:RGMaintext}, the critical temperature $T_c(U)$ has
the following functional form (see App.~\ref{app:RG})
\begin{equation}
T_c(U) = \Lambda \left (\frac{\Lambda}{T_c(\Lambda)}\right )^{-\frac{\Lambda}{U}}. \label{eq:Tc}
\end{equation}
The reference scale $T_c(\Lambda)$ depends on $J/U$, therefore both the magnitude of $U$ and of $J/U$ determine the scale of the instability. In contrast, rescaling of the running scale and coupling constants demonstrates that the phase diagram depends on the ratio $J/U$, only.

The RG equations were solved numerically using starting
values defined by Eqs.~\eqref{eq:bareinteraction} and Fermi velocities
$v_{\rm I} \approx 0.80 eV , v_{\rm II} \approx 0.86 eV , v_{\rm III}
\approx 0.57 eV$ as obtained from Fig.~\ref{fig:DispRel}, restricting
the values to the range $J/U\in [0,0.8]$.
We considered a large variety of $U/\bar v \in
[0.02,10]$, all leading to the same phase diagram as presented for
three exemplary values of $U/\bar v$ in
Fig.~\ref{fig:PhaseDiagram}. As we have mentioned above, the Fermi
velocities are non-universal and depend on the chosen microscopic
parameters. Therefore, a different set of Fermi velocities is explored
in Appendix~\ref{app:RG}, yielding similar results.

Our analysis identifies four different phases, as illustrated 
in {figures ~\ref{fig:PhaseDiagram} and  ~\ref{fig:SchematicRG}}.
The distinguishing characteristic of each
phase is the set of coupling constants that diverges and the signs of
the divergences. Near the phase boundaries, large finite values and
true divergencies are numerically indistinguishible, leading to
minor numerical uncertainties in Fig.~\ref{fig:PhaseDiagram}. In three
out of four phases, several coupling constants diverge at the same
scale preserving a finite, often universal, ratio. We derive these
fixed ratios analytically, they imply an enhanced
symmetry\cite{LinFisher1998} at the attractive fix point and a
connection to the integrable Gross-Neveu models discussed below. To
determine the physical meaning of the phases, we bosonize the degrees
of freedom
\begin{equation}
a^{R,L}_{{\alpha}, \sigma} \sim e^{i\sqrt{\pi} (\Phi_{{\alpha}, \sigma} \pm \Theta_{{\alpha},\sigma})}
\end{equation}
and perform a semiclassical strong coupling analysis near the instability. The latter allows to characterize gapful bosonic modes and to classify the operators displaying quasi-long range order. Following the lingo of 1D ladder systems, we use the notation ``C$m$S$n$'' for a phase with $m$ ($n$) massless bosons in the charge (spin) sector. Details on this procedure can be found in Appendix ~\ref{app:RG}. Here we summarize its outcome and discuss the phases in ascending order of $J/U$.

\subsection{Phase C3S2}
	For small Hund's coupling $J/U \lesssim 0.16$ only $\tilde c^{\sigma}_{\rm I, I}$ diverges towards negative infinity (attraction in the Cooper channel), while all other coupling constants remain featureless. This is due to the small starting value of $\tilde c^{\sigma}_{\rm I, I} = J$ which places the system close to a Cooper instability and in turn is due to the different orbital structure of left and right moving particles near $k_{\rm I}$. A spin gap is developed near $k_{\rm I}$ while all excitations near $k_{\rm II, III}$ remain gapless. Long-range correlations for singlet superconducting (SS) and also of charge density wave (CDW) order parameters occur in the C3S2 phase. When transformed back to the orbital and chain space, the superconducting order parameter takes the form
	\begin{equation}
	\Delta \sim \Delta_{\rm I} \Psi_{\rm I}^R(x) [\Psi_{\rm I}^{L}(x)]^T  = \frac{\Delta_{\rm I}}{4} 
	\left  (\begin{array}{cc}
	1 & -1 \\ 
	-1 & 1
	\end{array} \right )_\tau \otimes \left  (\begin{array}{cc}
	1 & -i \\ 
	i & 1
	\end{array} \right )_\gamma. \label{eq:Delta1}
	\end{equation}
	We have included indices $\tau$ and $\gamma$ in the matrix representation to clarify the direct product of chain and orbital spaces.
	Thus, the gap function contains a significant amount of orbital entanglement~\cite{GaoZhu2010,OngSchmalian2016} and has opposite sign along the rung and the steps of the ladder. As such it could be referred to as "d-wave", however, in 1D the notion of s-,d-, ... (p-,f-...) wave pairing is not well defined and we use the term even- (odd-) parity superconductivity instead. The gap function may be transformed to a real matrix by means of a $\pi/2$-rotation about the z-axis in orbital space.

\subsection{Phase C2S1a} At intermediate $0.16 \lesssim J/U \lesssim
0.41$ coupling constants involving only Fermi points $k_{\rm II}$
and $k_{\rm III}$ diverge, while those which involve Fermi point
$k_{\rm I}$ are unaffected.  The coupling constants scale as follows:
$\tilde c^\sigma_{\alpha\beta}, \tilde f^\rho_{\alpha\beta} \rightarrow +\infty$ and
$\tilde c^\rho_{\alpha\beta}, \tilde f^\sigma_{\alpha\beta} \rightarrow -\infty$ for
${\alpha, \beta} \in \lbrace \text{II,III}\rbrace$. For all starting values within
phase C2S1a, the divergence of diagonal spin components is
subdominant, such that $\tilde c_{{\alpha \alpha}}^\sigma/\tilde c_{\rm II,
III}^\sigma \rightarrow 0$, while the other coupling constants diverge
with a universal, finite ratio near the fix point. This corresponds to
the runaway flow in a certain, well defined direction of parameter
space, such that the flow becomes effectively one-dimensional. In the
Appendix, we expand the full RG equations about this ray and determine
the ratios of divergence, amongst others $\tilde c_{\rm II, III}^\rho
= - \tilde c_{\rm II, III}^\sigma/4$.

As compared to the three other phases, C2S1a displays repulsion in the Cooper channel. By means of the outlined semiclassical evaluation of the bosonized theory we find that the following bosonic modes are gapped
\begin{equation}
\Phi_{\rm II}^\rho - \Phi_{\rm III}^\rho, \; \Phi_{\rm II}^s - \Phi_{\rm III}^s,\; \Theta_{\rm II}^s +\Theta_{\rm III}^s,
\end{equation}
where $\Phi^{\rho,s}_{{\alpha}} = (\Phi_{{\alpha, }\uparrow} \pm \Phi_{{\alpha, }\downarrow})/\sqrt{2}$ are charge and spin modes, respectively. The refermionization of bosonic degrees of freedom in these three sectors yields a Gross Neveu model and highlights the emergent $SO(6) \sim SU(4)$ symmetry, see App.~\ref{app:RG}. Physically, the symmetry broken state corresponds to an orbitally ordered charge density wave. Returning to orbital and chain space, the order parameter is
\begin{equation}
O_{CDW}(x) = \left  (\begin{array}{cc}
	1 & 1 \\ 
	1 & 1
	\end{array} \right )_\tau \otimes\gamma_y  \otimes \mathbf 1_\sigma\; e^{-i(k_{\rm II} + k_{\rm III}-\pi)x}.
\label{eq:CDWSDWinC2S1a}
\end{equation}
	
\subsection{Phase C2S1b}	
	A rather narrow phase occurs for $0.42 \lesssim J/U \lesssim 0.53$. Again, coupling constants involving only Fermi points $k_{\rm II}$ and $k_{\rm III}$ diverge, while those which involve Fermi point $k_{\rm I}$ are featureless. As in the C2S1a phase, the ratios of the divergent coupling constants is universal and the effective one-dimensional RG flow is derived in Appendix \ref{app:RG}. The divergent quantities are $\tilde c_{\alpha\beta}^{\rho,\sigma} \rightarrow - \infty$ and $\tilde f^\rho_{\alpha\beta} \rightarrow + \infty$, while $\tilde f^\sigma_{\alpha\beta}/\tilde{c}_{{\alpha \alpha}}^\sigma \rightarrow -0$. Amongst the various finite ratios, we remark that $\tilde{c}_{\rm II, II}^\sigma/v_{\rm II} = \tilde{c}_{\rm III, III}^\sigma/v_{\rm III}$ and $\tilde{c}_{\rm II, III}^\rho = \tilde{c}_{\rm II, III}^\sigma/4$. Such a phase was discussed in detail in Ref.~\cite{LinFisher1997} and contains two massless charged and one massless spin mode while 
\begin{equation}
\Theta_{\rm II}^s, \; \Theta_{\rm III}^s,\; \Phi_{\rm II}^\rho -\Phi_{\rm III}^\rho,
\end{equation}
are gapped. Again, the connection to an $SO(6)$ Gross Neveu model with enlarged symmetry can be obtained by refermionization in the three sectors in which the bosons condense. Physically, this phase corresponds to an intraband superconductor with gaps $\Delta_{\rm II}, \Delta_{\rm III}$ on the Fermi points $k_{\rm II},k_{\rm III}$, in which $\Delta_{\rm II} \Delta_{\rm III} >0$. In orbital and chain space, the spin singlet gap function is the sum of the following contributions
	\begin{subequations}
	\begin{align}
	\Delta_{\rm II} \Psi_{\rm II}^R(x) [\Psi_{\rm II}^{L}(x)]^T  &= \frac{\Delta_{\rm II}}{2} 
	\left  (\begin{array}{cc}
	1 & 1 \\ 
	1 & 1
	\end{array} \right )_\tau \otimes \left  (\begin{array}{cc}
	1 & 0 \\ 
	0 & 0
	\end{array} \right )_\gamma , \\
	\Delta_{\rm III} \Psi_{\rm III}^R(x) [\Psi_{\rm III}^{L}(x)]^T  &= \frac{\Delta_{\rm III}}{2} 
	\left  (\begin{array}{cc}
	1 & 1 \\ 
	1 & 1
	\end{array} \right )_\tau \otimes \left  (\begin{array}{cc}
	0 & 0 \\ 
	0 & 1
	\end{array} \right )_\gamma.
	\end{align}\label{eq:Delta23}
	\end{subequations}
	
	This superconducting state relies on intraorbital pairing and has the same sign along and across the ladder.
	
\subsection{Phase C1S0}	
	Finally, for $0.53 \lesssim J/U $ coupling constants involving any Fermi surface diverge. The Cooper coupling constants $\tilde{c}_{\rm I,II}^{\sigma, \rho}, \tilde{c}_{\rm I,III}^{\sigma, \rho} \rightarrow + \infty $ while all other $\tilde{c}_{\alpha\beta}^{\sigma, \rho} \rightarrow - \infty$. As in the C2S1b phase, for non-equal ${\alpha \neq \beta}$ the relationship $4\tilde{c}_{{\alpha \beta}}^\rho /\tilde{c}_{{\alpha \beta}}^\sigma \rightarrow 1$ holds. In the forward scattering channel $\tilde{f}_{\alpha\beta}^\sigma/\tilde{c}_{\alpha\beta}^{\sigma, \rho} \rightarrow 0$, while $\tilde{f}_{\alpha\beta}^\rho/\tilde{c}_{\alpha\beta}^\sigma$ approaches a small constant value for ${\alpha \neq \beta}$. The C1S0 phase is the analog of the C2S1b phase, with the only difference that now all three Fermi points display an instability towards a symmetry broken state and that intraband Cooper couplings diverge independently $\tilde{c}_{\rm I, I}^\sigma/v_{\rm I} \neq \tilde{c}_{\rm II, II}^\sigma/v_{\rm II} \neq \tilde{c}_{\rm III, III}^\sigma/v_{\rm III}$. The semiclassical analysis of the bosonized theory predicts spin gaps for all three spin modes $\Theta_{\rm i}^s$ and charge gaps for the following two degrees of freedom. 
\begin{equation}
\Phi_{\rm I}^\rho- \Phi_{\rm II}^\rho,\; \Phi_{\rm I}^\rho -\Phi_{\rm III}^\rho.
\end{equation}
		This state represents a fully gapped superconductor with gaps $\Delta_{\rm I,II,III} $ and the charge modes lock in a manner such that 
	\begin{equation}
	\text{sign}(\Delta_{\rm I}\Delta_{\rm II}) = \text{sign}(\Delta_{\rm I}\Delta_{\rm III}) = -\text{sign}(\Delta_{\rm II}\Delta_{\rm III}). \label{eq:s++-}
	\end{equation}
	This follows from the positive signs of $\tilde{c}_{\rm I,III}^{\sigma,\rho}$ and $\tilde{c}_{\rm I,II}^{\sigma,\rho}$. Such a state may be called ``s$_{+--}$ '', its gap function is the sum of Eqs.~\eqref{eq:Delta1} and \eqref{eq:Delta23} with signs as imposed by Eq.~\eqref{eq:s++-}. The only massless bosonic mode is the overall phase of the superfluid.

\subsection{Discussion}
\label{sec:RG:Comparison}

In this section we include a discussion of the results, in particular of the superconducting phases. 

\subsubsection{RG flow}

We begin with an examination of the RG flow. In this context it is useful to introduce the Luttinger parameter matrices $ K^{\rho}_{\alpha\beta} \simeq \delta_{\alpha\beta} - 2\mathcal C_{\alpha\beta}^{\rho}, 
K^{\sigma}_{\alpha\beta} \simeq \delta_{\alpha\beta} +\mathcal C_{\alpha\beta}^{\sigma}/2$ with
\begin{equation}
\mathcal C_{\alpha\beta}^{{\rho, \sigma}} =\frac{1}{\pi(v_\alpha  + v_\beta )} \left (\begin{array}{ccc}
 \tilde c_{\rm I,I}^{{\rho, \sigma}} & \tilde f_{\rm I,II}^{{\rho, \sigma}} &\tilde f_{\rm I,III}^{{\rho, \sigma}} \\ 
\tilde  f_{\rm I,II}^{{\rho, \sigma}} & \tilde c_{\rm II,II}^{{\rho, \sigma}} &\tilde f_{\rm II,III}^{{\rho, \sigma}} \\ 
\tilde  f_{\rm I,III}^{{\rho, \sigma}} &  \tilde f_{\rm II,III}^{{\rho, \sigma}} &\tilde  c_{\rm III,III}^{{\rho, \sigma}}
\end{array} \right )_{ \alpha\beta}.
\end{equation}
The RG flow can be subdivided into two stages, see
Fig.~\ref{fig:SchematicRG}: In a first step, $K_{\alpha\beta}^\sigma$
renormalizes down towards $\delta_{\alpha\beta}$ with $K_{\alpha\beta}^\rho$ being
barely affected. Technically, this is due to terms of the standard
Cooper form $d \mathcal C_{\alpha\beta}^\sigma/d \ln(\Lambda/T) = -(\mathcal
C_{\alpha\beta}^\sigma)^2$ in the RG equations~\eqref{eq:RGequations}. This
state corresponds to a spinful {LL} with a non-interacting
spin sector, as is customary. Its quantum critical nature is crucial
in the present context as it represents a repulsive fixed point. Near
the fixed point the flows {diverge} and, in the second stage, the system
flows towards one of the four phases discussed above. By consequence,
the set of coupling constants $(\tilde c_{\alpha\beta}^{\rho, \sigma}, \tilde
f_{\alpha\beta}^{\rho, \sigma})$ in the infrared bears very little resemblance
with the bare high-energy parameters. The pattern of
interactions is completely reshuffled by many body effects. Notably,
$K_{{\alpha \alpha}}^\rho-1$ changes sign from intraband repulsion to intraband
attraction for ${\alpha} =$ II, III (${\alpha} =$ I,II,III) in the superconducting
phases C2S1b (C1S0) at lowest energies.

 When a trajectory approaches the quantum critical point very closely,
the flow slows down and $T_c$ shoots up, this is the origin of the
small $T_c$ in the C2S1 phases. Furthermore, since small starting
values of $U$ are closer to $K_{\alpha\beta}^\sigma = \delta_{\alpha\beta}$ this also
explains the dependence $T_c(U)$ as found in Eq.~\eqref{eq:Tc}. We
also note that the appearance of phases with gaps on a subset of the
Fermi points is rather generic in N-leg ladders \cite{LinFisher1997}.

More specific technical observations follow. The trajectories towards
C3S2 and C1S0 do not approach the QCP so closely leading to higher
$T_c$. In particular, the vanishing bare value $\tilde{c}_{\rm
I,I}^\sigma$ at $J \rightarrow 0$ places the system close to the
superconducting instability already. At larger $J$ the repulsive flow
near the QCP is driven by the RG equations~\eqref{eq:RGequations}
which imply that the initially vanishing $\tilde c_{\rm II,
III}^{\sigma, \rho}$ are increasing in magnitude due to finite $\tilde
f_{\rm II, III}^{\sigma, \rho}$.  

Finally, the divergence of $\tilde
c_{\rm II, III}^{\sigma, \rho}$ feeds back into the other channels,
which are small at intermediate scales.  This provides a mechanism to
explain the transition near $J \sim U/2$ between phases C2S1a, in
which $\tilde c_{\rm II,III}^\sigma \rightarrow +\infty$, and C2S1b,
in which $\tilde c_{\rm II,III}^\sigma \rightarrow -\infty$. Namely,
the RG equation for $\tilde c_{\rm II, III}^{\sigma}$ contains the
term $\tilde c_{\rm II, III}^{\sigma}[\tilde f^{\rho}_{\rm II, III}
-\tilde f_{\rm II, III}^{\sigma}/2] $. The square bracket changes sign
at $J = U/2$, see Eq.~\eqref{eq:bareinteraction}.  It is important to
stress that $\tilde{c}_{\rm II, III}^{\sigma,\rho}$ have vanishing
initial values, while the other interpocket Cooper interactions,
$\tilde{c}_{\rm I, II}^{\sigma,\rho}$ and $\tilde{c}_{\rm I,
III}^{\sigma,\rho}$ are repulsive (positive). Even after
renormalization, $\tilde{c}_{\rm I, II}^{\sigma,\rho}$ and
$\tilde{c}_{\rm I, III}^{\sigma,\rho}$ retain their positive sign,
which ultimately leads to the s$_{+--}$ superconductor in the C1S0
phase at largest Hund's coupling $J$.

\subsubsection{Physical implications and comparison to 2D}

We now turn the attention towards the physical implications of the
results.  Based on Fig.~\ref{fig:PhaseDiagram} b), in which $T_c$ is
estimated in Kelvin based on a bandwidth $\Lambda \sim 1eV$, we
conclude the following: First, none of the phases has an
experimentally relevant $T_c$ (above $1K$) in the controlled weak
coupling regime. We therefore extrapolate our analysis to larger
interaction amplitudes under the assumption that the RG flow is at
least qualitatively unchanged. Second, the phases with realistically
observable $T_c$ at intermediate coupling are C3S2 and C1S0, both of
which have long-range superconducting correlations. The orbital order
charge density wave C2S1a occurs at unrealistic energy scales, only.

It is interesting to
compare the present 1D RG for the ladder material with previous
studies of parquet RG~\cite{MaitiChubukov2010,ChubukovFernandes2016}
designed for materials with cylindrical Fermi surfaces. Both
approaches involve
weak coupling theory, and the basic form of RG equations,
Eq.~\eqref{eq:RGMaintext}, is the same, {and thus} 
Eq.~\eqref{eq:Tc} has an analogue in the 2D case. While the general
observation that many body effects completely reshuffle the pattern of
interaction constants persists to the higher dimensional systems, the
clear two stage RG as observed in
Figs.~\ref{fig:U08J01_RG}-\ref{fig:U08J06_RG} seems to be specific to
quasi 1D. {In contrast with our q1D study}, the 2D kinematics implies a
``parquet-to-ladder'' crossover scale given by $E_F \ll \Lambda$,
below which the general weak coupling form Eq.~\eqref{eq:RGMaintext}
(parquet) takes the simpler form $\beta_{\mu \nu \rho} = \delta_{\mu
\nu} \delta_{\nu \rho} \beta_\mu$ (ladder). Finally, an important
common observation valid both for q2D and q1D is that intraband
Coulomb interaction changes sign in the last stages of RG. Hence the
RG predicts superconducting pairing states with sign changes between
Fermi surfaces.

In view of this last point, it {is often assumed }
that RG resolves the ``Coulomb problem'': the 
question {of how} Coulomb repulsion {is overcome in the superconducting
state of iron pnictides and chalcogenides} in a wide variety of
different Fermi surface configurations, {without a significant
impact on the transition temperature}. {We }
recently investigated this {issue} in more
detail~\cite{KoenigColeman2017} and came to the conclusion that
generically RG is not sufficient to explain the 
{{robustness}}
of
superconductivity against the Coulomb repulsion. 
In the
present study the diverging Cooper attraction {{develops
at the end of the scaling trajecties}}
in channels with small or even vanishing bare couplings. 
{The energy scale at which this occurs is strongly dependent on
detailed microscopic interactions, suggesting
}
a {corresponding} dependence of $T_c$ on the microscopic details. 
Consequently, {although RG is able to account for the appearance of
pairing in a variety of different Fermi surface structures, } it does
not provide a natural explanation of the robustness of the
superconducting transition temperatures in iron-based superconductors,
and does not solve the Coulomb problem. {{This unsolved problem ,
which lies at the heart of the ubiquituous superconductivity in
the family of iron-based superconductors remains
an important challenge for the future.}}

\section{Half-filling: Umklapp scattering}
\label{sec:Umklapp}

The goal of this section is to qualitatively discuss umklapp
operators, Eq.~\eqref{eq:UmklappMainText}, by analyzing their scaling
dimension $d_u$. 
We
remind the reader that two electrons occupy each site at half filling,
cf. Fig.~\ref{fig:LadderOrbitals}, as a consequence umklapp scattering
involves six fermionic operators.  Due to the different spin structure
discussed in Sec.~\ref{sec:umklappMainText}, there is a total of
$2^7=128 $ umklapp terms. Their coupling constants are most
conveniently parametrized by seven complex numbers $A^\rho,
A_{\alpha}^s, B_{\alpha}^s$
\begin{equation}
\mathcal H_u =\text{Re}[A^\rho e^{i \sqrt{6\pi} \Theta_{\rm tot}^\rho}] \prod_{{\alpha} =\rm I}^{\rm III} \text{Re}[A_{\alpha}^s e^{i \sqrt{2\pi} \Theta_{{\alpha}}^s}+B_{\alpha}^s e^{i \sqrt{2\pi} \Phi_{{\alpha}}^s}]. \label{eq:UmklappDiscusssion}
\end{equation}
We introduced $\Theta_{\rm tot}^\rho = \sum_{\alpha} \Theta_{\alpha}^\rho/\sqrt{3}$
in the sector of total U(1) charge, the associated Luttinger parameter
is $K_{\rm tot}^\rho = \sum_{\alpha\beta} K_{\alpha\beta}^\rho/3$. We discuss the
scaling dimension of the umklapp operators in the vicinity of the five
fixed points presented in Fig.~\ref{fig:SchematicRG}, details are
relegated to Appendix~\ref{app:umklapp}. If umklapp scattering is RG
relevant, it can lock the bosonic field $\Theta_{\rm tot}^\rho$. Then
the system becomes an insulator with respect to the electromagnetic
U(1) charge and all superconducting phases are suppressed.

At the repulsive quantum critical point, the scaling dimensions of the
coupling constants are equal for all 128 umklapp terms and
\begin{equation}
d_u = (1-3 K_{\rm tot}^\rho)/2.
\end{equation}
Umklapp operators are relevant only at strong repulsion $K_{\rm tot}^\rho < 1/3$. We also remark that for an n-body interaction $d_u = 2 - n \stackrel{n = 3}{=} -1$ at the non-interacting fixed point.

In the phase C3S2 the phase $\Theta_{\rm I}^s$ condenses and therefore the product in Eq.~\eqref{eq:UmklappDiscusssion} involves $\alpha  = $II,III, only. Assuming $K_{\alpha\beta}^\sigma = \delta_{\alpha\beta}$ for $\alpha ,\beta  \in \lbrace {\rm II, III} \rbrace$ the scaling dimension becomes
\begin{equation}
d_u = 1-3 K_{\rm tot}^\rho/2.
\end{equation}
Mott localization occurs at intermediately strong coupling $K_{\rm tot}^\rho < 2/3$, only.

The phases C2S1a and C2S1b are characterized by a spin gap near both Fermi points $k_{\rm II}$ and $k_{\rm III}$. Effectively, the remaining umklapp terms have the form 
\begin{equation}
\mathcal H_u =\text{Re}[A^\rho e^{i \sqrt{6\pi} \Theta_{\rm tot}^\rho}]  \text{Re}[A_{\rm I}^s e^{i \sqrt{2\pi} \Theta_{\rm I}^s}+B_{\rm I}^s e^{i \sqrt{2\pi} \Phi_{\rm I}^s}]
\end{equation}
and (for $K_{\rm II}^\sigma = 1$) the scaling dimension
\begin{equation}
d_u = 3(1-K_{\rm tot}^\rho)/2.
\end{equation}
The transition occurs now at weak coupling, $K_{\rm tot}^\rho =1$. Note, however, that $K_{\alpha\beta}^\rho$ flows to attractive values near the fixed points. If Mott localization occurs C2S1a becomes a phase C1S0a in which the long range CDW correlations survive. Similarly, C2S1b becomes C1S0b, here superconducting correlations are suppressed but long range CDW correlations stemming from the Fermi point $k_{\rm I}$ persist.

Finally, the phase C1S0 has a spin gap at all Fermi momenta such that, effectively,  $\mathcal H_u =\text{Re}[A^\rho e^{i \sqrt{6\pi} \Theta_{\rm tot}^\rho}] $. The effective scaling dimension of the umklapp terms at the C1S0 fixed point is thus
\begin{equation}
d_u = 2-3 K_{\rm tot}^\rho/2,
\end{equation}
and Mott localization occurs at $K_{\rm tot}^\rho < 4/3$, i.e. formally even for attractive interaction. Then the system is fully gapped. In practice, the scaling dimension becomes relevant even before the C1S0 fixed point is reached. As we show explicitly in the Appendix, the dominant four umklapp operators have the form 
\begin{equation}
\mathcal H_u = g_u \text{Re}[A^\rho e^{i \sqrt{6\pi} \Theta_{\rm tot}^\rho}]\text{Re}[A^s e^{i \sqrt{6\pi} \Theta_{\rm rel}^s}] \label{eq:HC3S2Eff}
\end{equation}
with $\Theta_{\rm rel}^s = (\Theta_{\rm I}^s - \Theta_{\rm II}^s-\Theta_{\rm III}^s)/\sqrt{3}$. Even in the weak coupling limit, the scaling dimension of these operators change sign from irrelevant to relevant before reaching the C1S0 fixed point, see Fig.~\ref{fig:umklappDim} of the Appendix, and generate a C2S2 Mott insulating phase. 

In order to illustrate the appearance of a Mott phase we numerically solve Eq.~\eqref{eq:RGMaintext} along with $dg_u/dy = d_u (\lbrace g_\mu \rbrace) g_u$ in the parameter regime of strong Hund's coupling, see Fig.~\ref{fig:Mott}. For sufficiently large $U/\Lambda$ (e.g. $U/\Lambda \gtrsim 1.6$ for $J = 0.6 U$) $d_u>0$ even at the initial stage of RG. The divergence of the coupling constant is then dominated by the repulsive quantum critical fixed leading to
 \begin{equation}
g_u(y) \approx g_u(0) e^{[1 - 3K_{\rm tot}^\rho(y = 0)] y/2}. \label{eq:UmklappDivergence}
 \end{equation} 
In Fig.~\ref{fig:Mott}, we employ a working definition of the Mott activation gap $T_{\rm Mott}$ by means of the scale $y_{\rm Mott} = \ln (\Lambda/T_{\rm Mott})$ at which $g_u(y_{\rm Mott}) = (2\pi \Lambda)$ for starting value $g_u(0) = U^2/(2\pi \Lambda)$. 
For a density $n$ away from the density of half filling $n_0$, the
divergence of $g_u$ is cut at an energy scale $\bar v \vert n - n_0
\vert$ leading to $T_{\rm Mott}(n) = \sqrt{T_{\rm Mott}(n_0)^2- (\bar
v \vert n - n_0 \vert)^2}$. For clear illustration we chose $(\bar v
\vert n - n_0 \vert/\Lambda)^2 = 0.3$, i.e. a rather large value, in
Fig.~\ref{fig:Mott}. We note that  using $\Lambda \simeq 1 eV$, the phase boundaries of \ref{fig:Mott} occur at unrealistically low temperatures, which is a direct consequence of our weak coupling treatment. However, we conjecture that the qualitative outcome of our controlled calculations still holds in the strong coupling limit.

All in all, we conclude that Mott physics can be important for the present itinerant model. Umklapp scattering at half filling may be RG relevant near the attractive fixed points even though it is strongly irrelevant at the non-interacting limit.

\begin{figure}
\includegraphics[scale=.6]{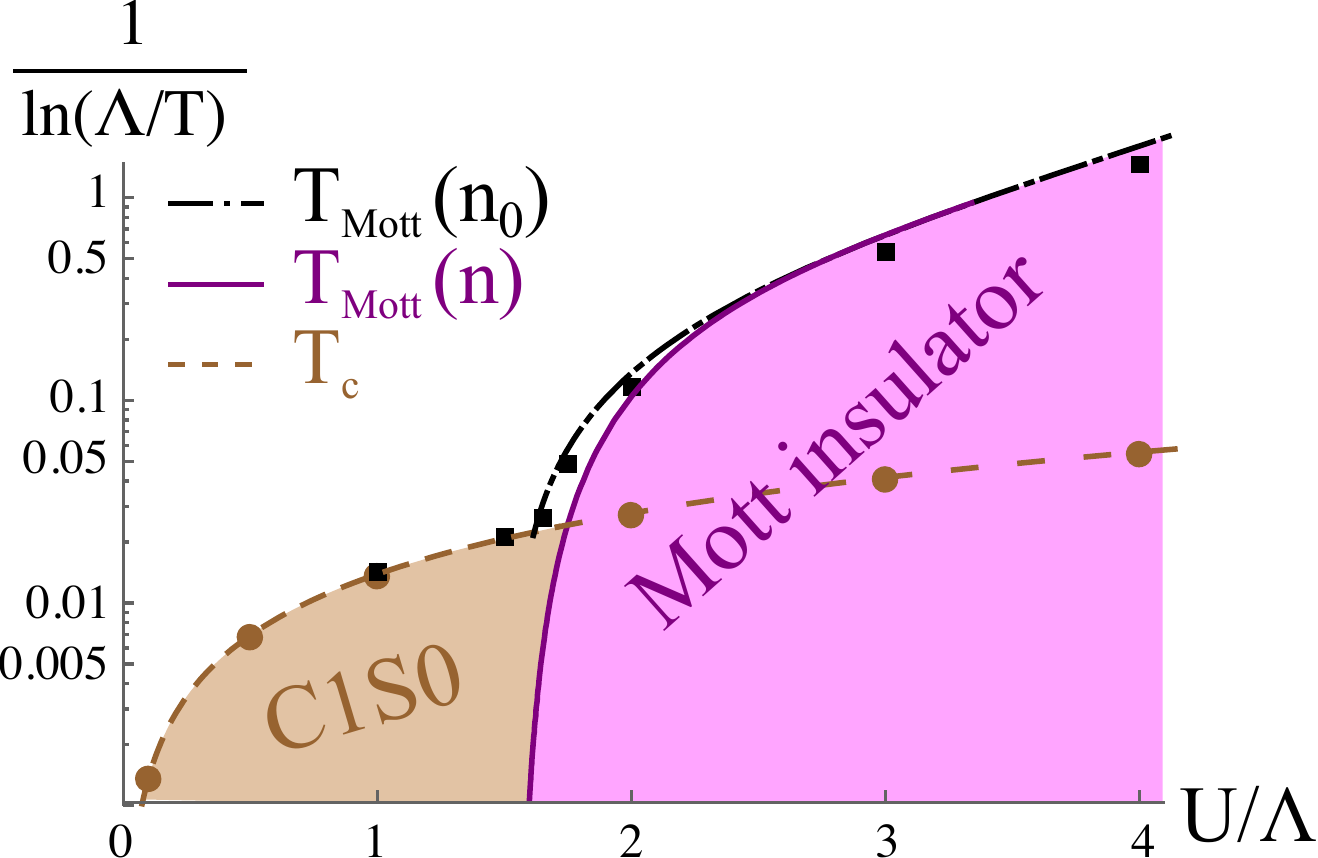}

\caption{Phase diagram at $J = 0.6 U$, taking umklapp scattering into
account. The quantity $U/\Lambda = U \tilde a/\bar v$ on the x-axis
decreases with pressure, while the  y-axis variable 
$1/\ln(\Lambda/T)$ increases with temperature. 
At large $U/\Lambda$ and
perfect commensuration, the superconducting $T_c$ (numerical solution:
brown dots; Eq.~\eqref{eq:Tc}: dashed brown) always lies below the 
Mott localization temperature $T_{\rm Mott} (n_0)$ defined
heuristically by $g_u(y_{\rm Mott}) = (2\pi \Lambda)$ (numerical
solution: black squares; approximate solution,
Eq.~\eqref{eq:UmklappDivergence}, valid for $U/\Lambda \gtrsim 1.6$:
black dotdashed). Away from half filling the Mott phase boundary is
modified by the finite energy scale associated with doping (purple, here $(\bar v \vert n - n_0
\vert/\Lambda)^2 = 0.3$) giving rise to a superconductor-insulator
transition.}  \label{fig:Mott}
\end{figure}

\section{Summary and outlook}

In summary, we have investigated a single ladder for the quasi 1D iron
based superconductor BaFe$_2$S$_3$ on the basis of the tightbinding
model as suggested in Refs.~\cite{AritaSuzuki2015, PatelDagotto2016}
and a simple onsite Hubbard and Hund interactions. 
The long-wavelength low-temperature physics was studied using an RG analysis of excitations close to the three pairs of Fermi points. 

In the case of incommensurate filling the weak coupling RG analysis
yields four phases depending on the ratio of Hund $J$ to Hubbard $U$
interactions, see Fig.~\ref{fig:PhaseDiagram}.  We have shown
analytically and checked numerically that the absolute value of the
interaction $ U$ at given $J/U$ does not affect the ground state, but
$T_c$ increases rapidly as a function of $U$.  In particular, a fully
gapped superconductor is stabilized at sufficiently large ratio $J/U
\gtrsim 0.53 $. The intraband pairing gaps have signs $+, -, -$ on the
three pairs of Fermi points and the critical temperature $T_c$ is
estimated to be of the order of 10 K for intermediately strong
coupling.  {This} theory provides a way to understand the conducting
high pressure part of the experimental phase
diagram~\cite{YamauchiOhgushi2015}. {In order to account} for the
observation of a Mott phase at low pressure {we furthermore
investigated} umklapp scattering at half filling. {In our three band
model, umklapp processes are represented by three-body interactions
and are hence irrelevant at weak coupling. However, we have 
shown that umklapp scattering does become relevant near some of
the strong coupling fixed points.  In this circumstance a
charge gap develops, giving rise to a correlated
insulator.} Near commensurate filling these results indicate a
Mott insulator-superconductor quantum phase transition (see
Fig.~\ref{fig:Mott}) into a Hund's-driven superconducting
phase at intermediate repulsion. This corroborates DMRG studies on very
similar models~\cite{PatelDagotto2016,PatelDagotto2017}.

One of the fascinating aspects of our model, is that it is 
able to realize 
several different ground states which develop a dynamically
enhanced symmetry, each characterized by different universal
fixed-point ratios of the coupling constants. In the real
3D material the 1D renormalization group flows  will be 
cut-off by interchain hopping, which is a
relevant perturbation. In this spirit, the present study provides insight
into the predeliction towards certain superconducting and magnetic states
that the real material inherits from its 1D building blocks.

There are various lessons that 
we have learnt that are relevant to the wider 
theoretical study of iron based
superconductors. First, we have seen that the orbital structure of
wave functions near the Fermi surface plays a crucial role in the {
formation of the order parameters}. Second, the interorbital and
orbital selective pairings are ubiquitous as soon as the orbital
structure of the wave functions is taken into account. Third, even in
weak coupling theories, the Hund's coupling has a dramatic
impact. Finally, although RG studies enable us to understand 
the appearance of pairing in a wide variety of q1D Fermi surface
structures, this pairing still requires a channel with a weak bare Coulomb
interaction, and a  generic mechanism which
accounts for the  weak bare repulsion in the iron-based
superconductors is still needed}. 

\section{Acknowledgements}

We thank Y. Komijani and P.-Y. Chang for useful discussions.
E. K\"{o}nig and P. Coleman were supported by the U.S. Department of Energy, Basic
Energy Sciences, grant number DE-FG02-99ER45790.
A. M. Tsvelik was supported by the U.S. Department of Energy (DOE),  Division  of Condensed Matter Physics and  Materials Science, under Contract No. \text{DE-SC0012704}

\appendix

\section{Microscopic model}
\label{app:Micro}

\subsection{Tight binding model}
\label{app:tightbinding}
In this Appendix we present further details on the  tight binding model~\cite{PatelDagotto2016}, which we use to obtain input parameters for the numerical integration of RG equations.

We consider a ladder as shown in Fig.~\ref{fig:LadderOrbitals}. Our approach
employs the following notation for the $2\times 2 \times 2 = 8$ degrees of
freedom at each site: a) Spin is represented by Pauli matrices
$\boldsymbol{\sigma}$. Spin eigenvalues are $\uparrow, \downarrow$.
b) Pauli matrices in chain space are $\boldsymbol{\tau}$. We denote
the chains by $1,2$ (upper, lower chain). Due to mirror symmetry along
the chain direction, $\tilde \tau = \pm 1$ is a good quantum number
which corresponds to parity eigenstates $\frac{\ket{1} + \tilde \tau
\ket{2}}{\sqrt{2}}$.  c) Pauli matrices in orbital space are denoted
by $\boldsymbol{\gamma}$. The two orbitals are called $a, b$.  The
tight binding Hamiltonian contains nearest and next nearest rung hopping,
\begin{align}
H_{\rm kin} &=  \sum_{{j} \sigma} \Bigg \lbrace  [d_{j}^+ t_Z \mathbf 1_\tau d_{{j} + 1} + h.c. ]+ [d_{j}^+ t_{2Z} \mathbf 1_\tau d_{{j} + 2} + h.c.] \notag \\
&+[d_{j}^+ t_{Z+Y} \tau_1 d_{{j} + 1} + h.c.]+[d_{j}^+ t_{2Z+Y} \tau_1 d_{{j} + 2} + h.c. ]\notag \\
&+d^+_{j} t_Y \tau_1 d_{j} + \Delta d^+_{j} \gamma_3 d_{j} - \mu d^+_{j} d_{j} \Bigg \rbrace.
\end{align}
We note that $t_\mu$ with $\mu = Z,Y,Z+Y,2Z,2Z+Y$ are $2\times2$ matrices in orbital space and that their off-diagonal parts are antisymmetric and that we returned to the coordinate system $(X,Y,Z)$ to make contact with Ref.~\cite{PatelDagotto2016}. We omitted the spin index for notational simplicity, following Ref.~\cite{PatelDagotto2016} $t_Y$ is a diagonal matrix.

We introduce the Fourier transform $d_{j} = \frac{1}{N} \sum_k e^{ik
{j}} d_k$ and perform a transformation in chain space into the basis
of bonding/antibonding states. We define
\begin{equation}
d_{k} = \left (\begin{array}{c}
d_{k,\tilde \tau = +1} \\ 
d_{k,\tilde \tau = -1}
\end{array} \right ) = \frac{\tau_1 + \tau_3}{\sqrt{2}} \left (\begin{array}{c}
c_{k,1} \\ 
c_{k,2}
\end{array} \right ),
\end{equation}
where spin and orbital quantum numbers have been suppressed for
convenience. 
This leads to the following  result
\begin{subequations}
\begin{equation}
H_{\rm kin} = \sum_{\sigma = \uparrow\downarrow, \tilde \tau = \pm 1} \int (dk) d^+_{k, \sigma, \tilde\tau} \left [\sum_{i = 0,2,3} h_{\tilde \tau}^{(i)}\gamma_i\right ] d_{k, \sigma, \tilde \tau}
\end{equation}
where $d_{k,\sigma, {\tilde \tau}}$ are two spinors in orbital space,
$\gamma_0 = \mathbf 1_\gamma$ and we have introduced 
\begin{align}
h_{{\tilde \tau}}^{(0)} &= 2 \cos(k) [t_Z^{(0)} + {\tilde \tau} t_{Z+Y}^{(0)}] + 2 \cos(2k) [t_{2Z}^{(0)} + {\tilde \tau} t_{2Z+Y}^{(0)}] \notag \\
&- \mu + {\tilde \tau} t_Y^{(0)} \\
h_{{\tilde \tau}}^{(2)} &= 2 i\sin(k) [t_Z^{(2)} + {\tilde \tau} t_{Z+Y}^{(2)}] + 2 i\sin(2k) [t_{2Z}^{(2)} + {\tilde \tau} t_{2Z+Y}^{(2)}] \\
h_{{\tilde \tau}}^{(3)} &= 2 \cos(k) [t_Z^{(3)} + {\tilde \tau} t_{Z+Y}^{(3)}] + 2 \cos(2k) [t_{2Z}^{(3)} + {\tilde \tau} t_{2Z+Y}^{(3)}] \notag \\
& + \Delta + {\tilde \tau} t_Y^{(3)}.
\end{align}
We have also introduced $t_\mu^{(i)} = \text{tr}[t_\mu \gamma_i]/2$
with $\mu = Z,Y,Z+Y, 2Z, 2Z+Y$ and $i = 0,2,3$. In this notation,
$h_{{\tilde \tau}}^{(2)}$ is real and the time reversal symmetry $\left
[\sum_{i = 0,2,3} h_{\tilde \tau}^{(i)}\gamma_i\right ]^T = \left
[\sum_{i = 0,2,3} h_{\tilde \tau}^{(i)}\gamma_i\right ]_{k \rightarrow
-k}$ is apparent.
\end{subequations}
The spectrum of the tight binding Hamiltonian is then
\begin{equation}
\epsilon_{{\tilde \tau}, \gamma} (k) = h_{\tilde \tau}^{(0)} + \gamma \sqrt{(h_{\tilde \tau}^{(2)})^2 +(h_{\tilde \tau}^{(3)})^2}. \label{eq:DispRel}
\end{equation}
The plot of the spectrum for the parameters given in Eq. (3) of Ref.~\cite{PatelDagotto2016} is given in Fig.~\ref{fig:DispRel}. 

\subsection{Interaction terms} \label{app:Interaction} Here  we provide
more details about the interaction terms. It is useful to disentangle
interorbital from intraorbital contributions in the Hubbard
interaction, even though we set $\tilde U = U$ in the main text. In
the following ${\cal P}_{\gamma, \tau, \sigma}$ are projectors onto a
given orbital, chain, and spin respectively.
\begin{align}
H_{U} &= U \sum_{j} \sum_{\gamma = a,b} \sum_{ \tau = 1,2} (d_{{j},\uparrow}^\dagger {\cal P}_\gamma {\cal P}_{ \tau} d_{{j},\uparrow})(d_{{j},\downarrow}^\dagger{\cal P}_\gamma {\cal P}_{ \tau} d_{{j},\downarrow}) \label{eq:HubbardIntraorbital}\\
H_{\tilde U} &= \tilde U \sum_{j}  \sum_{ \tau = 1,2} (d_{{j}}^\dagger {\cal P}_a {\cal P}_{ \tau} \mathbf 1_\sigma d_{{j}})(d_{{j}}^\dagger {\cal P}_b {\cal P}_{ \tau} \mathbf 1_\sigma d_{{j}})\label{eq:HubbardInterorbital} 
\end{align}

As a next step, we carry out the long-wavelength expansion,
Eq.~\eqref{eq:LowEnExp}, into the interaction terms of the
Hamiltonian. As usual, the overall momentum conservation may be
preserved in three different manners, corresponding to direct,
exchange and Cooper channels.

\paragraph{Intraorbital Hubbard interaction}
We begin by rewriting the intraorbital Hubbard interaction, Eq.~\eqref{eq:HubbardIntraorbital} in terms of low-energy modes. 

\begin{subequations}
In the density channel we then obtain
\begin{align}
H_U^{(0)} &= \frac{U}{8} \int dx \sum_{\rm {\alpha} = II, III} \left (\sum_{r} (a_{\rm I, \uparrow}^{r, \dagger}a_{\rm I, \uparrow}^{r} + 2 a_{{\alpha, }\uparrow}^{r, \dagger}a_{{\alpha, }\uparrow}^{r})\right ) \notag \\
&\times \left (\sum_{r'} (a_{\rm I, \downarrow}^{r', \dagger}a_{\rm I, \downarrow}^{r'} + 2 a_{{\alpha, }\downarrow}^{r', \dagger}a_{{\alpha, }\downarrow}^{r'})\right ),
\end{align}
while in the exchange channel the interaction takes the form 
\begin{align}
H_U^{(X)} &= -\frac{U}{8} \int dx \sum_{\rm {\alpha} = II, III} \left (\sum_{r} (a_{\rm I, \uparrow}^{r, \dagger}a_{\rm I, \downarrow}^{r} + 2 a_{{\alpha, }\uparrow}^{r, \dagger}a_{{\alpha, }\downarrow}^{r})\right )\notag \\
&\times \left (\sum_{r'} (a_{\rm I, \downarrow}^{r', \dagger}a_{\rm I, \uparrow}^{r'} + 2 a_{{\alpha, }\downarrow}^{r', \dagger}a_{{\alpha, }\uparrow}^{r'})\right ).
\end{align}
Finally, in the Cooper channel we obtain
\begin{align}
H_U^{(C)} &= \frac{U}{8} \int dx \sum_{\rm {\alpha} = II, III} \left (\sum_{r} (a_{\rm I, \uparrow}^{r, \dagger}a_{\rm I, \downarrow}^{-r,\dagger} + 2 a_{{\alpha, }\uparrow}^{r, \dagger}a_{{\alpha, }\downarrow}^{-r,\dagger})\right )\notag \\
&\times \left (\sum_{r'} (a_{\rm I, \downarrow}^{r'}a_{\rm I, \uparrow}^{-r'} + 2 a_{{\alpha, }\downarrow}^{r'}a_{{\alpha, }\uparrow}^{-r'})\right ).
\end{align}
\end{subequations}
\paragraph{Interorbital Hubbard interaction}
Next we look at the interorbital Hubbard term, Eq.~\eqref{eq:HubbardInterorbital}.
\begin{subequations}
In the density channel we obtain
\begin{align}
H_{\tilde U}^{(0)} &= \frac{\tilde U}{8} \int dx \left (\sum_{\sigma,r} (a_{\rm I, \sigma}^{r, \dagger}a_{\rm I, \sigma}^{r} + 2 a_{\rm II, \sigma}^{r, \dagger}a_{\rm II, \sigma}^{r})\right )\notag \\
& \times \left (\sum_{\sigma',r'} (a_{\rm I, \sigma'}^{r', \dagger}a_{\rm I, \sigma'}^{r'} + 2 a_{\rm III, \sigma'}^{r', \dagger}a_{\rm III, \sigma'}^{r'})\right ),
\end{align}
while in the exchange term, 
\begin{equation}
H_{\tilde U}^{(X)} = - \frac{\tilde U}{8}\int dx \sum_{\sigma, \sigma'} \left (\sum_{r} r a_{\rm I, \sigma}^{r, \dagger}a_{\rm I, \sigma'}^{r} \right )\left (\sum_{r'} r' a_{\rm I, \sigma'}^{r', \dagger}a_{\rm I, \sigma}^{r'}\right ).
\end{equation}
In the Cooper channel
\begin{equation}
H_{\tilde U}^{(C)} = - \frac{\tilde U}{8}\int dx \sum_{\sigma, \sigma'}\left (\sum_{r} r a_{\rm I, \sigma}^{r, \dagger}a_{\rm I, \sigma'}^{-r, \dagger} \right )\left (\sum_{r'} r' a_{\rm I, \sigma'}^{r'}a_{\rm I, \sigma}^{-r'}\right )
\end{equation}
only operators from the Fermipoint $k_{\rm I}$ are involved. 
\end{subequations}

\paragraph{Hund's coupling.} The treatment of Hund's coupling is 
analagous to the treatment of the interorbital Hubbard interaction.

\begin{subequations}
In the density channel we obtain
\begin{align}
H_{J}^{(0)} &= -\frac{J}{8} \int dx \left (\sum_{r} (a_{\rm I}^{r, \dagger}\vec \sigma a_{\rm I}^{r} + 2 a_{\rm II}^{r, \dagger}\vec \sigma a_{\rm II}^{r})\right )\notag \\
& \times \left (\sum_{r'} (a_{\rm I}^{r', \dagger}\vec \sigma a_{\rm I}^{r'} + 2 a_{\rm III}^{r', \dagger}\vec \sigma a_{\rm III}^{r'})\right ),
\end{align}
while in the exchange,
\begin{align}
H_{J}^{(X)} &= \frac{ J}{8}\int dx \sum_{\sigma, \sigma'} \left (\sum_{r} r \vec \sigma a_{\rm I}^{r} a_{\rm I}^{r, \dagger}\right )_{\sigma \sigma'} \notag \\
& \times \left (\sum_{r'} r' \vec \sigma a_{\rm I}^{r'}a_{\rm I}^{r', \dagger}\right )_{\sigma ' , \sigma}
\end{align}
and Cooper channel,
\begin{align}
H_{J}^{(C)} &= \frac{J}{8}\int dx \sum_{\sigma, \sigma'}\sum_{\tilde \sigma, \tilde \sigma'}(\vec \sigma^T)_{\sigma, \tilde \sigma}(\vec \sigma)_{\sigma', \tilde \sigma'}\left (\sum_{r} r a_{\rm I, \tilde \sigma'}^{r, \dagger}a_{\rm I, \sigma}^{-r, \dagger} \right )\notag \\
& \times \left (\sum_{r'} r' a_{\rm I, \tilde \sigma'}^{r'}a_{\rm I, \sigma'}^{-r'}\right ).
\end{align}
only operators from the Fermipoint $k_{\rm I}$ are involved. 
\end{subequations}

At $U = \tilde U$, the interaction terms presented in this Appendix may be rearranged in the form of scalar and vector current densities and ultimately yield Eq.~\eqref{eq:bareinteraction} of the main text. In contrast, if $\tilde U = U - \tilde J$ additions to the coupling constants with the following form arise
\begin{subequations}
\begin{eqnarray}
 \delta \tilde f_{\alpha\beta}^\rho &=& - \frac{\tilde J}{4}\left (\begin{array}{ccc}
 0 & 1 & 1 \\ 
 1 & 0 & 2 \\ 
 1 & 2 & 0
 \end{array} \right )_{\alpha\beta}, \\
 \delta  \tilde f_{\alpha\beta}^\sigma &=& 0 ,\\
 \delta \tilde c_{\alpha\beta}^\rho &=& -  \frac{\tilde J}{4} \left (\begin{array}{ccc}
 3 & 0 & 0 \\ 
 0 & 0 & 0 \\ 
 0 & 0 & 0
 \end{array} \right ) _{\alpha\beta} , \\
 \delta  \tilde c_{\alpha\beta}^\sigma &=& \tilde J \left (\begin{array}{ccc}
 1 & 0 & 0 \\ 
 0 & 0 & 0 \\ 
 0 & 0 & 0
 \end{array} \right ) _{\alpha\beta}{.}
\end{eqnarray}
\label{eq:bareinteractioncorrections}
\end{subequations}

\section{RG flow - incommensurate filling}
\label{app:RG}

In this appendix we collect representative numerical solutions of the RG
equations for each of the four phases presented in
Fig.~\ref{fig:PhaseDiagram} of the main text. We also present
some technical details from the analysis of the phases, begining
with the 
$\tilde J = 0$ phases,  in increasing order of $J/U$. 
In Fig.~\ref{fig:PhaseDiagram_v108_v208_v306} we show the
phase diagram obtained for a different set of Fermi velocities to those
used in Fig.~\ref{fig:PhaseDiagram}, showing how the phase
boundaries shift with velocities. We have left the discussion of
finite $\tilde J$ for appendix \ref{app:NonEqualHubbard}

\begin{figure}
\includegraphics[scale=.7]{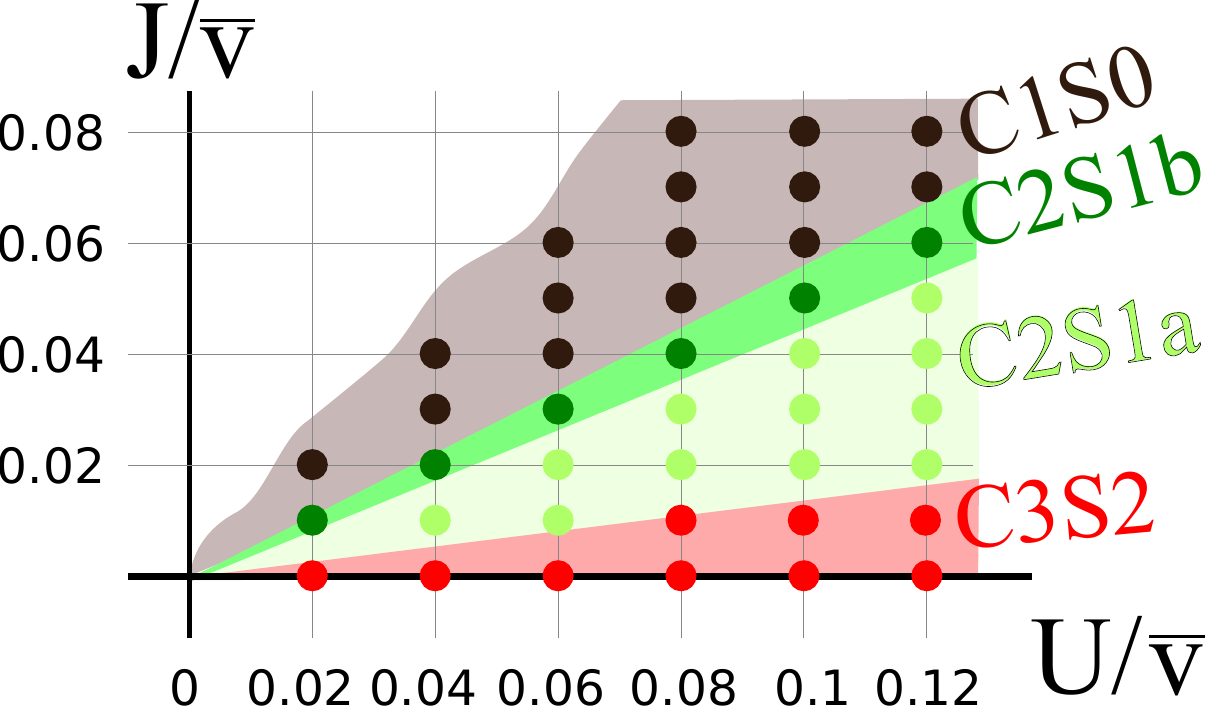}
\includegraphics[scale=.7]{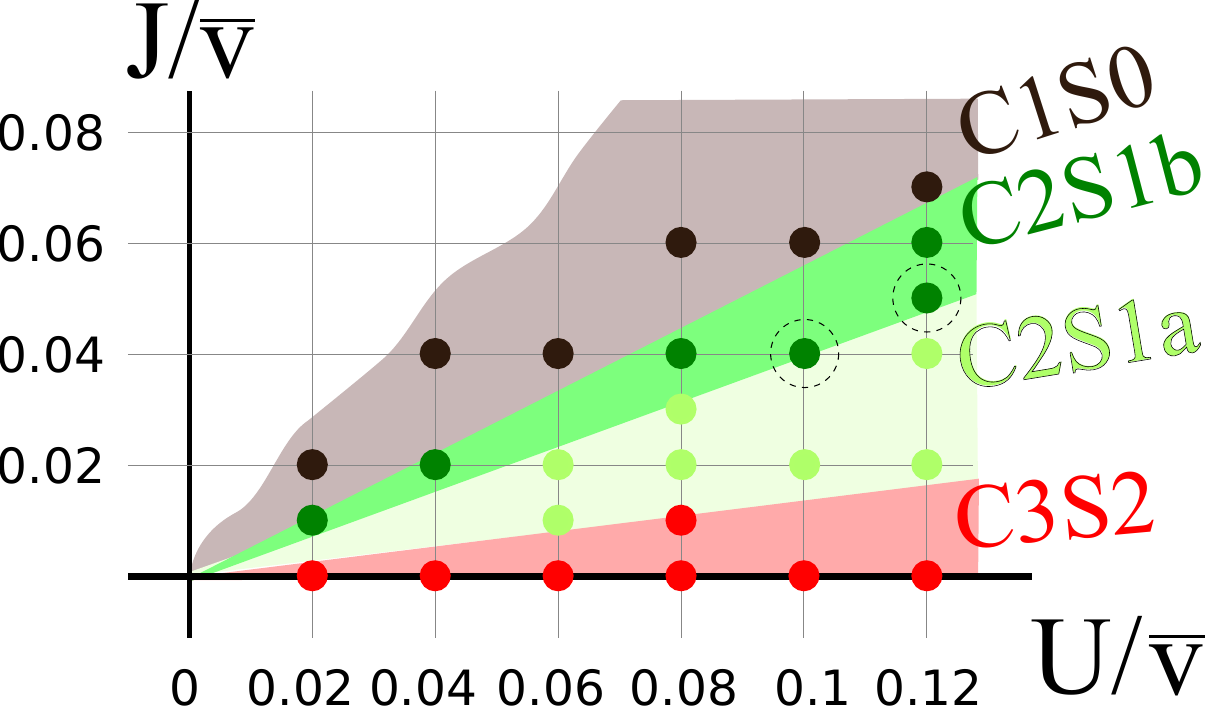}
\caption{Phase diagram for small Hubbard $U$ 
and Hund's $J \leq U$, for the set of Fermi velocities used in
Fig.~\ref{fig:DispRel} (top) and $v_{\rm I} = v_{\rm II} = 0.8 eV,
v_{\rm III} = 0.6 eV$ (bottom). In the second case, the system at $U
= 0.1 \bar v, J = 0.04 \bar v$ and $U = 0.12 \bar v, J = 0.05 \bar v$
flows to the phase C2S1b (dashed circles), thus the boundary between
C2S1b and C2S1a is shifted downwards.}
\label{fig:PhaseDiagram_v108_v208_v306}
\end{figure}

\subsection{RG equations}

The RG equations for an N-leg ladder without umklapp scattering were derived in Ref.~\cite{LinFisher1997}
\begin{subequations}
\begin{eqnarray}
\dot f_{\alpha\beta}^\rho &=& (c^\rho_{\alpha\beta})^2 + \frac{3}{16} (c_{\alpha\beta}^\sigma)^2,\\
\dot f_{\alpha\beta}^\sigma &=& -(f_{\alpha\beta}^\sigma)^2 + 2 c_{\alpha\beta}^\rho c_{\alpha\beta}^\sigma  - \frac{1}{2} (c_{\alpha\beta}^\sigma)^2,\\
\dot c_{\alpha\beta}^\rho &=& - \sum_{\gamma} \lbrace \alpha_{{\alpha \beta, \gamma}} (c_{{\alpha \gamma}}^\rho c_{{\gamma \beta}}^\rho + \frac{3}{16} c_{{\alpha \gamma}}^\sigma c_{{\gamma \beta}}^\sigma)) \rbrace, \notag \\
&&+(c_{\alpha\beta}^\rho h_{\alpha\beta}^\rho +\frac{3}{16} c_{\alpha\beta}^\sigma h_{\alpha\beta}^\sigma)\\
\dot c_{\alpha\beta}^\sigma &=& - \sum_{\gamma} \lbrace \alpha_{{\alpha \beta, \gamma}} (c_{{\alpha \gamma}}^\rho c_{{\gamma \beta}}^\sigma + c_{{\alpha \gamma}}^\sigma c_{{\gamma \beta}}^\rho + \frac{1}{2} c_{{\alpha \gamma}}^\sigma c_{{\gamma \beta}}^\sigma)) \rbrace, \notag \\
&&+(c_{\alpha\beta}^\rho h_{\alpha\beta}^\sigma + c_{\alpha\beta}^\sigma h_{\alpha\beta}^\rho -\frac{1}{2} c_{\alpha\beta}^\sigma h_{\alpha\beta}^\sigma)
\end{eqnarray}
\label{eq:RGequations}
\end{subequations}

Parameters without the tilde are defined by $f_{\alpha\beta}^\rho = {\tilde f_{\alpha\beta}}/{\pi(v_{\alpha} + v_{\beta})}$ etc., the ratio $\alpha_{{\alpha \beta, \gamma}} = (v_{\alpha} + v_{\gamma})(v_{\beta}+ v_{\gamma})/[2 v_{\gamma} (v_{\alpha} + v_{\beta})]$ and $h_{\alpha\beta}^{\rho, \sigma}  = 2f_{\alpha\beta}^{\rho, \sigma} + \delta_{\alpha\beta} c_{\alpha\beta}^{\rho, \sigma}$. The dot indicates the derivative with respect to the running scale, $\dot c = d c/d \ln(L/\tilde a)$, where $\tilde a$ is the UV length scale.

\subsection{Relationship $T_c(U)$}
Here, we derive Eq.~\eqref{eq:Tc} of the main text. It is useful to employ the schematic representation Eq.~\eqref{eq:RGMaintext} of the RG equations. If the coupling constants $g_\mu(y)$ obey the RG equations, so do $\bar g_\mu (y)  = g_\mu(y/U)/U$. Let $g_\mu$ have an instability at $y_c(g_\mu^0)$, where $y_c(g_\mu^0)$ is an unknown function of the bare values $g_\mu^0$. Clearly, the instability of $\bar g_\mu$ occurs at $y_c(g_\mu^0/U)$ and in view of the relationship between $ g_\mu (y)$ and $\bar g_\mu (y)$ it follows that
\begin{equation}
{y_c(g_\mu^0/U)} =U  y_c (g_\mu^0).
\end{equation}
Using $y_c = \ln (\Lambda/T_c)$ Eq.~\eqref{eq:Tc} of the main text follows.

\subsection{Analysis of RG flow}
\paragraph{C3S2 phase}

The phase of smallest $J/U$ is characterized by the lone divergence of
$c_{\rm I,I}^\sigma \rightarrow - \infty$, while all other
coupling constants remain featureless (see
Fig.~\ref{fig:U08J01_RG}). Expanding the set of RG equations in
powers of $c_{\rm I,I}^\sigma$ one we confirm that the flow of
$c_{\rm I,I}^\sigma$ decouples from all other RG equations and
diverges as $\dot{c}_{\rm I,I}^\sigma = -(c_{\rm I,I}^\sigma)^2$.

\begin{figure}
\includegraphics[scale=.41]{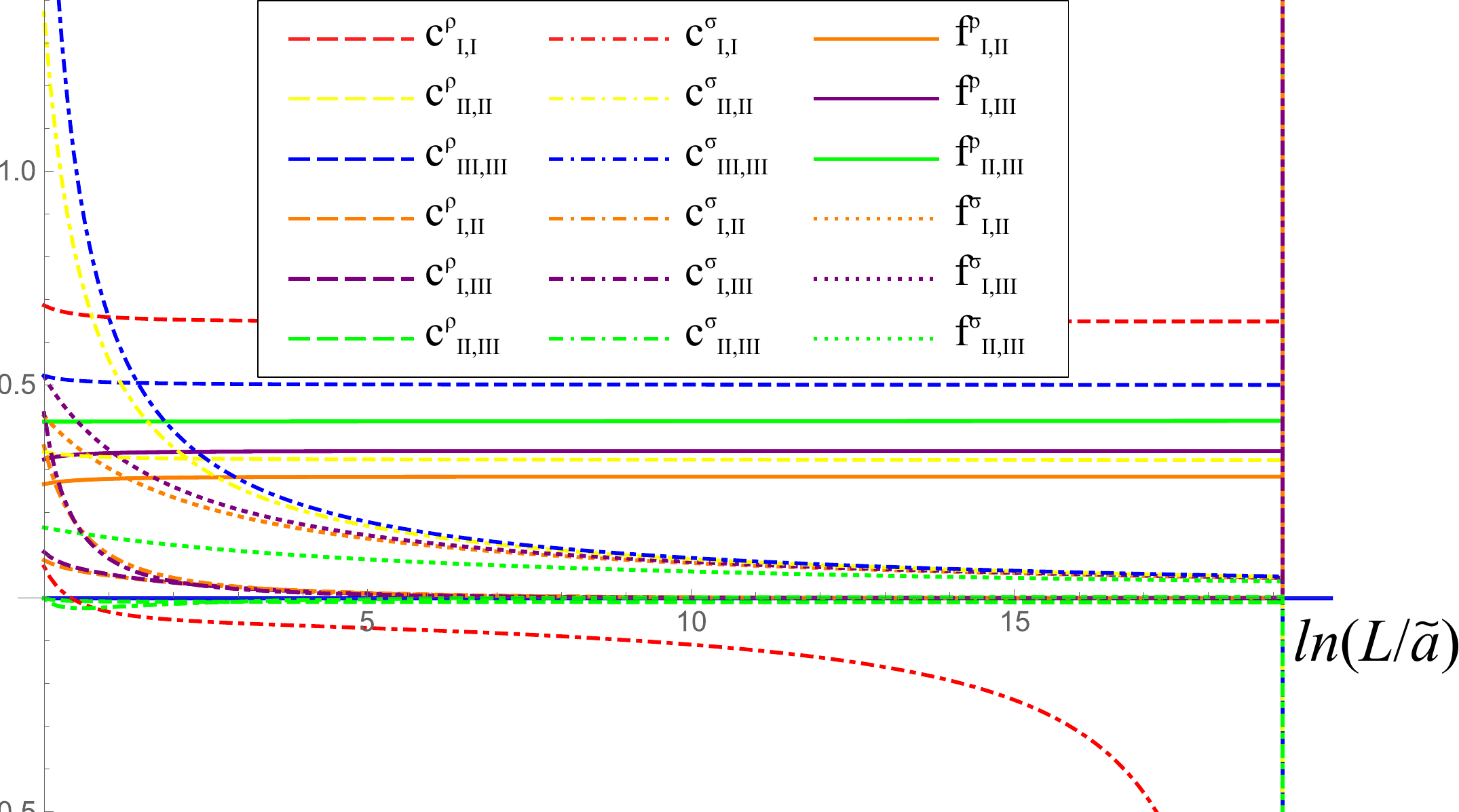} 
\caption{Numerical integration of RG equations for starting values determined by $U/\bar v = 5$ and $J/U = 0.1$. In this case, the system flows to the C3S2 phase, which is characterized by divergent $c_{\rm I,I}^\sigma$.}
\label{fig:U08J01_RG}
\end{figure}

\paragraph{C2S1a phase.}

The phase of second smallest $J/U$, plotted in light green in
Fig.~\ref{fig:PhaseDiagram}, is characterized by a divergence of
several coupling constants, while the ratio to $c_{\rm II,
III}^\sigma \rightarrow + \infty$ is fixed throughout the phase
\begin{subequations}
\begin{eqnarray}
c_{\rm II, II}^\rho = c_{\rm III, III}^\rho &=& - \frac{R_1}{8} c_{\rm II, III}^\sigma \\
c_{\rm II, III}^\rho &=& - \frac{1}{4} c_{\rm II, III}^\sigma \\
f_{\rm II, III}^\rho &=& \frac{1}{8} R_2 c_{\rm II, III}^\sigma \\
f_{\rm II, III}^\sigma &=& - R_3 c_{\rm II, III}^\sigma.
\end{eqnarray}
\label{eq:CouplingConstantsLightGreen}
\end{subequations}
The intraband, spin-spin interactions are equal and have a subdominant
divergence $c_{\rm II, II}^\sigma = c_{\rm III, III}^\sigma
\rightarrow +\infty$ but are small in comparison to the coupling
constants of Eq.~\eqref{eq:CouplingConstantsLightGreen}, i.e. $c_{\rm
II, II}^\sigma/c_{\rm II, III}^\sigma \rightarrow 0$. All other
coupling constants remain small (see Fig.~\ref{fig:U08J02_RG}). The
ansatz \eqref{eq:CouplingConstantsLightGreen}, when introduced into
the full RG equations, proves to be consistent provided ($\zeta = 1/2
+ v_{\rm II}/[4v_{\rm III}] + v_{\rm III}/[4v_{\rm II}]$ is
non-universal)
\begin{subequations}
\begin{eqnarray}
R_1 &=& \zeta R_2\\
R_2 &=& \frac{2 R_3}{R_3^2 + 1}\\
R_3 &=& \sqrt{\sqrt{\frac{8  + \zeta^2}{4}} - \frac{\zeta}{2}} 
\end{eqnarray}
\end{subequations}
For $v_{\rm II} \rightarrow v_{\rm III}$ all $R_{1,2,3}$ approach
unity. The set of RG equations reduces to a single equation for one of
the six parameters, $\dot c_{\rm II, III}^\sigma = 2 (c_{\rm II,
III}^\sigma)^2/R_2$.

\begin{figure}
\includegraphics[scale=.4]{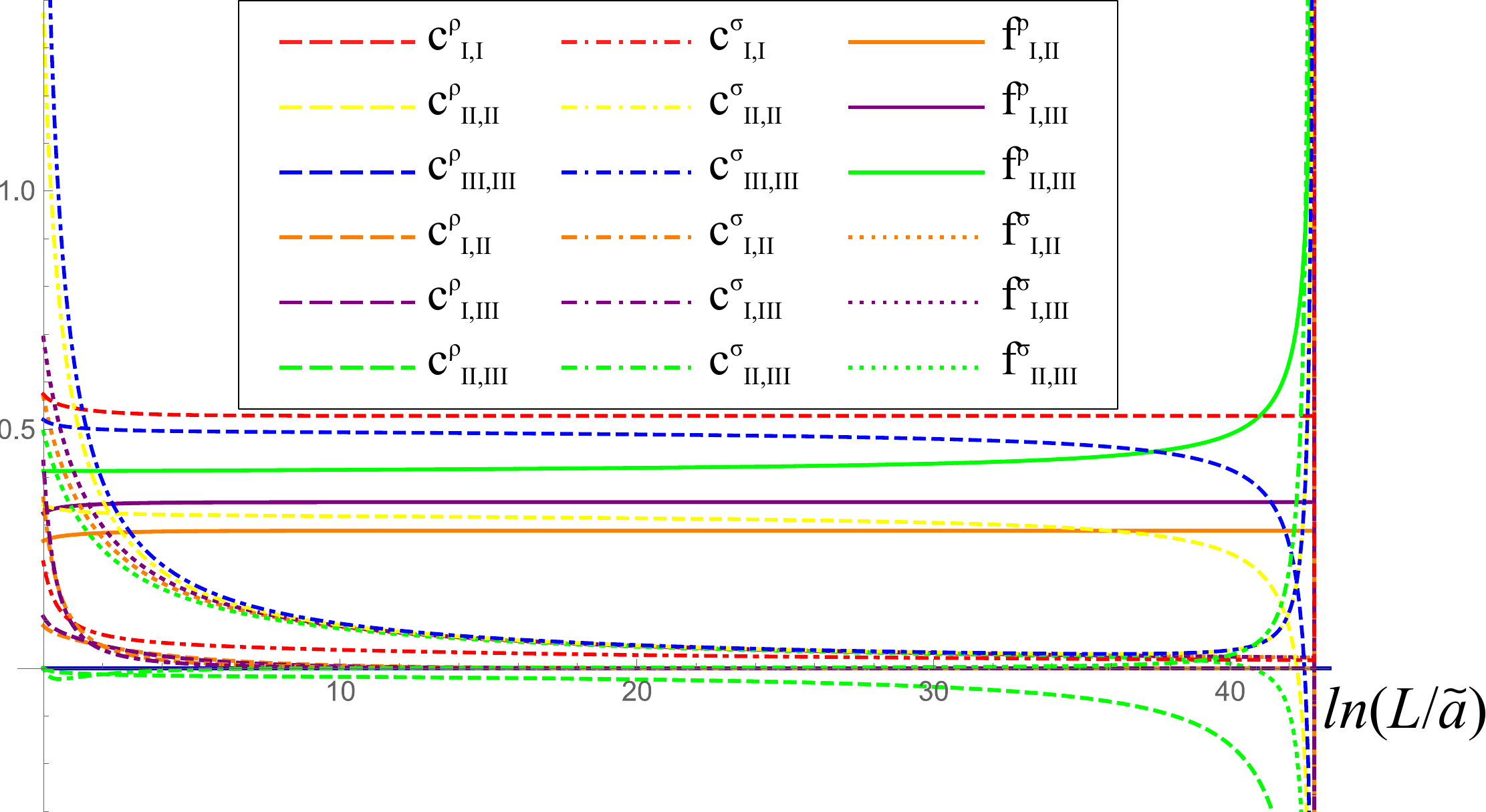} 
\caption{Numerical integration of RG equations for starting values determined by $U/\bar v = 5$ and $J/U = 0.3$. In this case, the system flows to the C2S1a phase.}
\label{fig:U08J02_RG}
\end{figure}

\paragraph{C2S1b phase.}

The C2S1b phase is characterized by the following diverging running
coupling constants where the sign of $c_{\rm II, III}^\sigma
\rightarrow - \infty$ constitutes the key  difference with the
phase previously discussed.

\begin{subequations}
\begin{eqnarray}
c_{\rm II, II}^\rho = c_{\rm III, III}^\rho &=& \frac{R_1}{8} c_{\rm II, III}^\sigma \\
c_{\rm II, II}^\sigma = c_{\rm III, III}^\sigma  &=& R_2 c_{\rm II, III}^\sigma \\
c_{\rm II, III}^\rho &=& \frac{1}{4} c_{\rm II, III}^\sigma \\
f_{\rm II, III}^\rho &=& -\frac{R_3}{8} c_{\rm II, III}^\sigma.
\end{eqnarray}
\end{subequations}
Though the coupling constants $f_{\rm II, III}^\sigma$, 
increase near criticality, they remain relatively small $f_{\rm II, III}^\sigma/c_{\rm II, III}^\sigma \rightarrow 0$.
Again, the ratios $R_{1,2,3}$ are constant throughout the phase and given in terms of $\alpha$ by the following functions which approach unity as $\zeta \rightarrow 1$. 

\begin{subequations}
\begin{eqnarray}
R_1 &=& \zeta R_3\\
R_2 &=& \sqrt{\sqrt{\frac{8 \zeta^2+1}{4}} - \frac{1}{2}} \\
R_3 &=& \frac{2 R_2}{R_2^2 + \zeta}
\end{eqnarray}
\end{subequations}

The coupled divergences are captured by the single RG equation $\dot c_{\rm II, III}^\sigma = -2 (c_{\rm II, III}^\sigma)^2/R_3$.

\begin{figure}
\includegraphics[scale=.4]{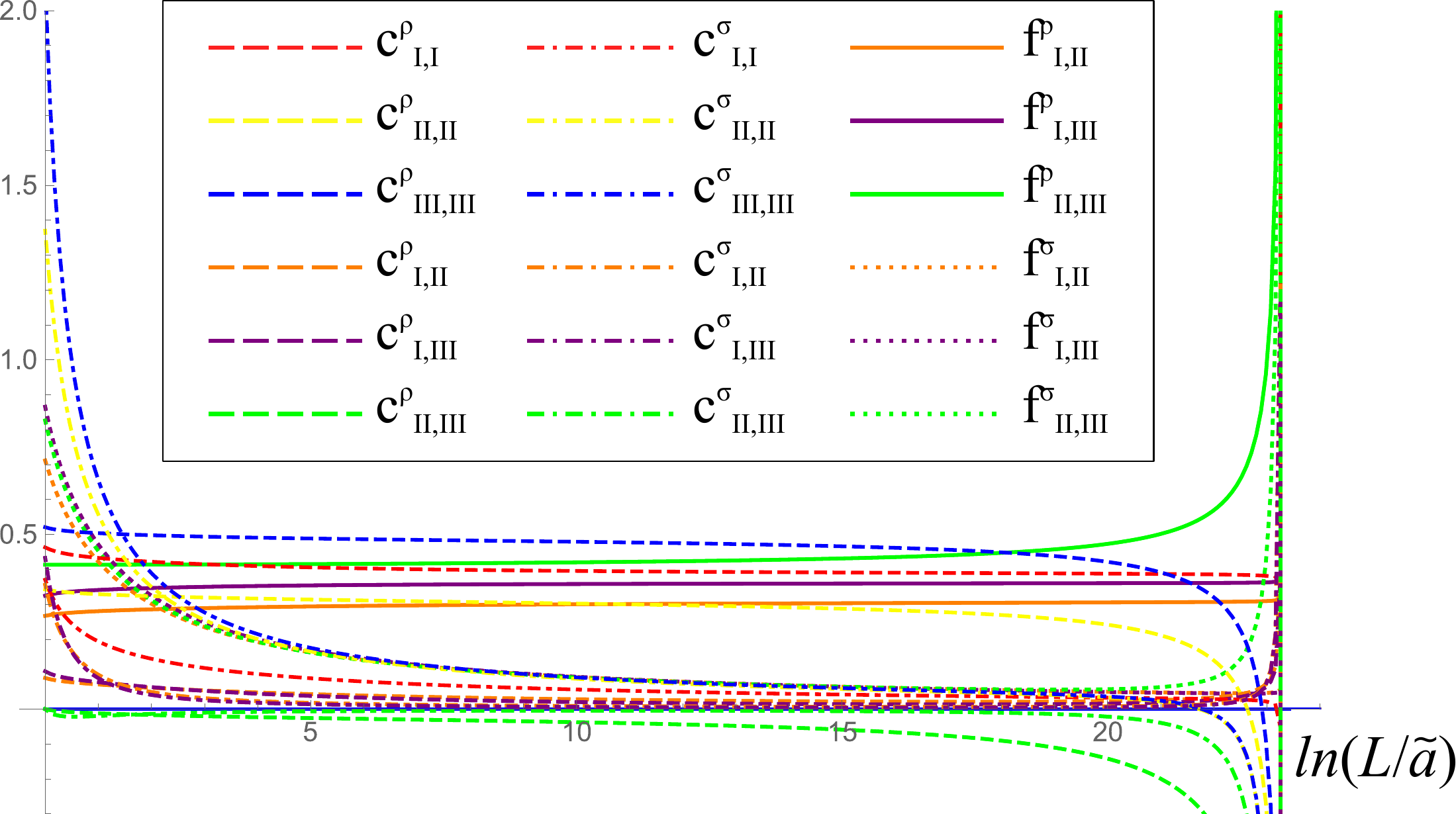} 
\caption{Numerical integration of RG equations for starting values determined by $U/\bar v = 5$ and $J/U = 0.5$. In this case, the system flows to the C2S1b phase. Note that, as compared to Fig.~\ref{fig:U08J02_RG}, $c_{\rm III,III}^\sigma$ flows towards negative infinity.}
\label{fig:U08J04_RG}
\end{figure}
\paragraph{C1S0 phase}

Finally, at the largest $J$ we considered, coupling constants
involving any of the three Fermi points diverge. As in the C2S1b phase
and in Ref.~\cite{LinFisher1997}, the ratio $c_{\alpha\beta}^\rho =
c_{\alpha\beta}^\sigma/4$ for ${\alpha \neq \beta}$ is preserved and
$f_{\alpha\beta}^\sigma \rightarrow 0$. In contrast with the previous
discussion, the divergence of the intra Fermi point couplings
$c_{{\alpha \alpha}}$ is generically $i$ dependent. The sign of the
diverging coupling constants is
\begin{subequations}
\begin{eqnarray}
c_{{\alpha \alpha}}^\sigma &\rightarrow & - \infty\\
c_{{\alpha \alpha}}^\rho &\rightarrow & -\infty \\
c_{\rm I, II}^\sigma = 4c_{\rm I, II}^\rho &\rightarrow &  +\infty\\
c_{\rm II, III}^\sigma = 4c_{\rm II, III}^\rho&\rightarrow & - \infty\\
c_{\rm I, III}^\sigma= 4c_{\rm I, III}^\rho &\rightarrow & + \infty \\
f_{\alpha\beta}^\rho & \rightarrow & +\infty 
\end{eqnarray}
\label{eq:CouplingConstantsBrown}
\end{subequations}
Once again, $f_{\alpha\beta}^\sigma >0$ is small as compared to the
couplings discussed in Eq.~\eqref{eq:CouplingConstantsBrown} and
formally $f_{\alpha\beta}^\sigma \rightarrow 0$ at the fixed point.

\begin{figure}
\includegraphics[scale=.4]{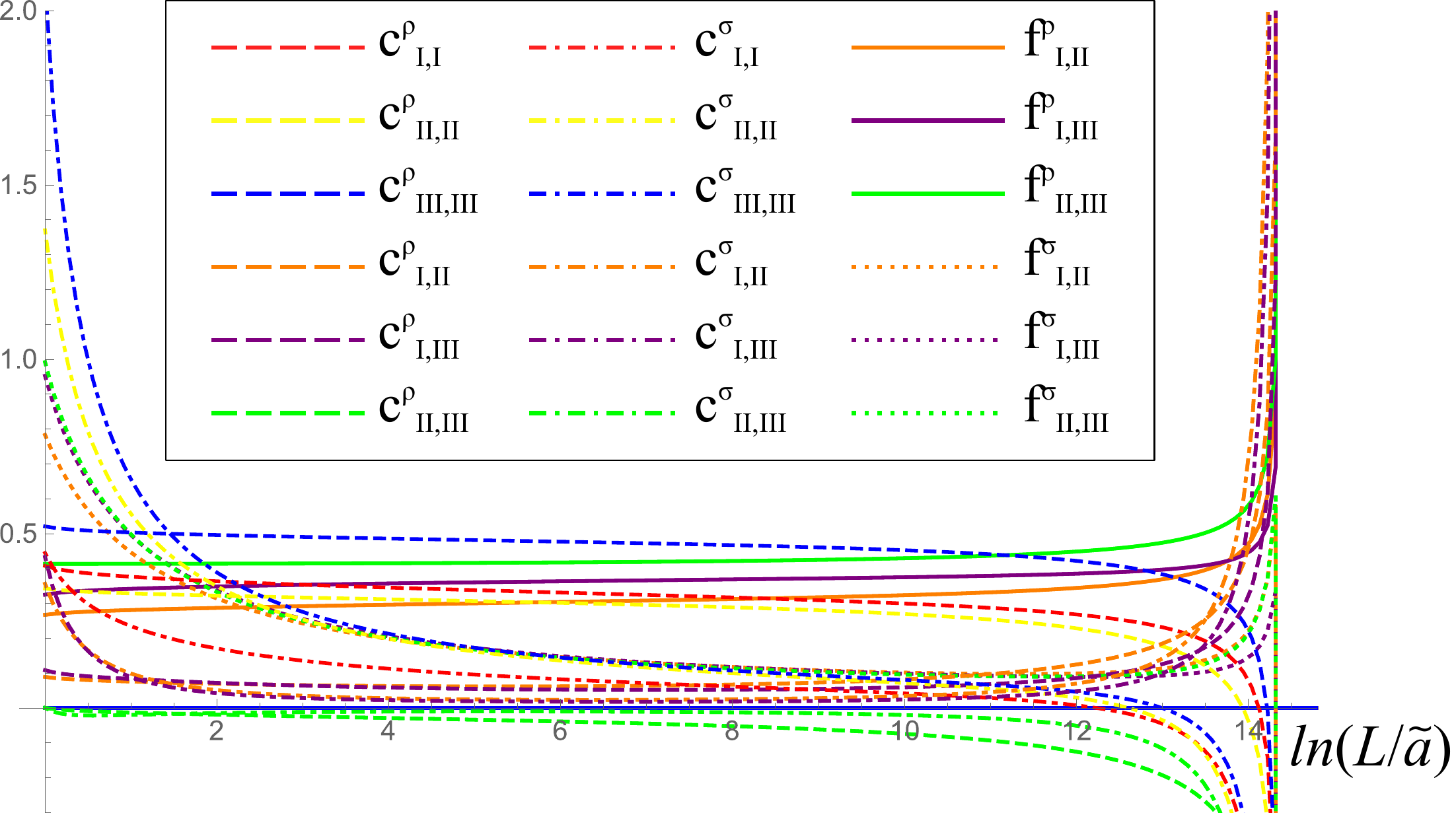} 
\caption{Numerical integration of RG equations for starting values determined by $U/\bar v = 5$ and $J/U = 0.6$. In this case, the system flows to the C1S0 phase.}
\label{fig:U08J06_RG}
\end{figure}

\subsection{Physical meaning of the different phases}

In the previous section we analyzed the RG flow and distinguished
four different phases, characterized by the divergence of 
four different sets
of coupling constants at a critical, exponentially large
length scale $L_*$. Here, we study the physical implications of these
four cases.  In this analysis we follow Ref.~\cite{LinFisher1997} and
bosonize the theory near $L_*$ subsequently using a
semiclassical analysis.

\subsubsection{Bosonization Dictionary}
For the bosonization we use the dictionary 
\begin{subequations}
\begin{equation}
a^{r}_{{\alpha},\sigma}(x) = \frac{1}{\sqrt{2\pi a}} \eta_{{\alpha} \sigma} e^{i \sqrt{4\pi} \phi_{{\alpha},\sigma}^r(x)}
\end{equation}
with the following identities
\begin{eqnarray}
\left  \lbrace \eta_{{\alpha},\sigma}, \eta_{{\alpha}',\sigma'} \right \rbrace &=& 2 \delta_{{\alpha \alpha}'} \delta_{\sigma, \sigma'} \\
\left [ \phi_{{\alpha},\sigma}^r (x), \phi_{{\alpha}',\sigma'}^{r'} (x') \right ] &=& \frac{i r}{4} \text{sign}(x - x') \delta_{rr'} \delta_{{\alpha \alpha}'} \delta_{\sigma, \sigma'} \notag \\
&+& \frac{i r}{4} \delta_{r,-r'} \delta_{{\alpha \alpha}'} \delta_{\sigma, \sigma'} .
\end{eqnarray}
\end{subequations}
The length scale $a_*$ is the UV cut-off of the renormalized
theory. Since we have assumed the same Majorana Klein factor for
creation and annihilation operators, normal ordering must be imposed
prior to bosonization in order to preserve the consistency of
signs. We assume that any operator under consideration contains an
even number of fermionic operators from each Fermipoint
${\beta}$. Thus, in the Klein factor Hilbert space, each operator
involving Fermi point ${\beta}$ contains one of the following 4
operators at least once:
\begin{equation}
\eta_{{\beta},\uparrow} \eta_{{\beta},\uparrow} = 1; \eta_{{\beta},\downarrow}\eta_{{\beta},\downarrow} = 1; \eta_{{\beta},\uparrow}\eta_{{\beta},\downarrow} \equiv g_{\beta};\eta_{{\beta},\downarrow}\eta_{{\beta},\uparrow} \equiv - g_{\beta}.  
\end{equation}
The algebra of $g_{\beta}$ is $g_{\beta}^2 = -1$ and $ [ g_{\gamma}, g_{\beta}  ] = 0$, a representation of this algebra is
\begin{equation}
g_{\beta} = i ,\quad \forall j.
\end{equation}

We introduce density and displacement fields
\begin{eqnarray}
\phi_{{\alpha},\sigma}^R &=& \frac{\Phi_{{\alpha},\sigma} + \Theta_{{\alpha},\sigma}}{2}, \\
\phi_{{\alpha},\sigma}^L &=& \frac{\Phi_{{\alpha},\sigma} - \Theta_{{\alpha},\sigma}}{2} .
\end{eqnarray}
Spin charge separation is accounted for by the parametrization
\begin{eqnarray}
\Phi_{\alpha}^{\rho} &=& \frac{\Phi_{{\alpha, }\uparrow} +\Phi_{{\alpha, }\downarrow}}{\sqrt{2}} \\
\Phi_{\alpha}^{s} &=& \frac{\Phi_{{\alpha, }\uparrow} -\Phi_{{\alpha, }\downarrow}}{\sqrt{2}} 
\end{eqnarray}
and analogously for $\Theta$ variables. Using our bosonization convention, $\Phi_{\alpha}^\rho$ turns out to be proportional to the collective phase for a superconducting ground state. 

\subsubsection{Operators under consideration.}
The phases under consideration are of the superconducting and charge density wave (CDW) /spin density wave (SDW) type. We study correlation functions of CDW, SDW$_z$ (z-component of the SDW order parameter), singlet superconductivity (SS) and the z-component of triplet superconductivity (TS$_z$). The phases under consideration regard the following intraband operators,
\begin{subequations}
\begin{align}
O_{\rm CDW}^{({\alpha})} &= (a_{{\alpha},\uparrow}^{R, \dagger}a_{{\alpha},\uparrow}^{L} + a_{{\alpha},\downarrow}^{R, \dagger}a_{{\alpha},\downarrow}^{L} ) e^{- i2k_{\alpha} x}\notag \\
&\sim \frac{1 }{(\pi a)} e^{- i \sqrt{2\pi} \Theta_{\alpha}^\rho} \cos(\sqrt{2\pi} \Theta_{\alpha}^s)e^{- i2k_{\alpha} x},\\
O_{\rm SDW_z}^{({\alpha})} &=(a_{{\alpha},\uparrow}^{R, \dagger}a_{{\alpha},\uparrow}^{L} - a_{{\alpha},\downarrow}^{R, \dagger}a_{{\alpha},\downarrow}^{L} ) e^{- i2k_{\alpha} x}\notag \\
&\sim \frac{1}{(\pi a)} e^{- i \sqrt{2\pi} \Theta_{\alpha}^\rho} \sin(\sqrt{2\pi} \Theta_{\alpha}^s)e^{- i2k_{\alpha} x},\\
O_{\rm SS}^{({\alpha})} &= (a_{{\alpha},\uparrow}^{R, \dagger}a_{{\alpha},\downarrow}^{L, \dagger} - a_{{\alpha},\downarrow}^{R, \dagger}a_{{\alpha},\uparrow}^{L, \dagger} ) \notag \\
&\sim \frac{1}{(\pi a)} e^{- i \sqrt{2\pi} \Phi_{\alpha}^\rho} \cos(\sqrt{2\pi} \Theta_{\alpha}^s),\\
O_{\rm TS_z}^{({\alpha})} &= (a_{{\alpha},\uparrow}^{R, \dagger}a_{{\alpha},\downarrow}^{L, \dagger} +a_{{\alpha},\downarrow}^{R, \dagger}a_{{\alpha},\uparrow}^{L, \dagger}) \notag \\
&\sim \frac{1}{(\pi a)} e^{- i \sqrt{2\pi} \Phi_{\alpha}^\rho} \sin(\sqrt{2\pi} \Theta_{\alpha}^s),
\end{align}
\begin{widetext}
as well as the following interband operators in the particle-hole channels
\begin{align}
O_{\rm CDW}^{(\alpha\beta)} &= \frac{1}{2}(a_{{\alpha},\uparrow}^{R, \dagger}a_{{\beta},\uparrow}^{L} + a_{{\alpha},\downarrow}^{R, \dagger}a_{{\beta},\downarrow}^{L}  ) e^{-i (k_{\alpha}+k_{\beta}) x} + {\alpha} \leftrightarrow {\beta} \notag \\
&\sim  e^{- i (k_{\alpha} + k_{\beta})x}\frac{e^{- i \sqrt{\pi} \Theta_{\alpha\beta}^{\rho +}}}{\pi a} \left (\cos(\sqrt{\pi} \Phi_{\alpha\beta}^{\rho -}) \cos(\sqrt{\pi } \Theta_{\alpha\beta}^{s +}) \sin(\sqrt{\pi } \Phi_{\alpha\beta}^{s -})  - i \sin(\sqrt{\pi } \Phi_{\alpha\beta}^{\rho -}) \sin(\sqrt{\pi } \Theta_{\alpha\beta}^{s +}) \cos(\sqrt{\pi }\Phi_{\alpha\beta}^{s -}) \right ) \\
O_{\rm SDW_z}^{(\alpha\beta)} &=\frac{1}{2}(a_{{\alpha},\uparrow}^{R, \dagger}a_{{\beta},\uparrow}^{L} - a_{{\alpha},\downarrow}^{R, \dagger}a_{{\beta},\downarrow}^{L}  ) e^{-i (k_{\alpha}+k_{\beta}) x}+ {\alpha} \leftrightarrow {\beta}\notag \\
& \sim e^{- i (k_{\alpha} + k_{\beta})x}\frac{e^{- i \sqrt{\pi} \Theta_{\alpha\beta}^{\rho +}}}{\pi a} \left (\cos(\sqrt{\pi } \Phi_{\alpha\beta}^{\rho -}) \sin(\sqrt{\pi } \Theta_{\alpha\beta}^{s +}) \sin(\sqrt{\pi } \Phi_{\alpha\beta}^{s -})  - i \sin(\sqrt{\pi } \Phi_{\alpha\beta}^{\rho -}) \cos(\sqrt{\pi } \Theta_{\alpha\beta}^{s +}) \cos(\sqrt{\pi } \Phi_{\alpha\beta}^{s -}) \right )\\
O_{\rm CDW}^{[\alpha\beta]} &= \frac{1}{2}(a_{{\alpha},\uparrow}^{R, \dagger}a_{{\beta},\uparrow}^{L} + a_{{\alpha},\downarrow}^{R, \dagger}a_{{\beta},\downarrow}^{L} ) e^{-i(k_{\alpha}+k_{\beta}) x} - {\alpha} \leftrightarrow {\beta} \notag \\
&\sim  e^{- i (k_{\alpha} + k_{\beta})x}\frac{e^{- i \sqrt{\pi} \Theta_{\alpha\beta}^{\rho +}}}{\pi a}\left (\cos(\sqrt{\pi } \Phi_{\alpha\beta}^{\rho -}) \sin(\sqrt{\pi } \Theta_{\alpha\beta}^{s +}) \cos(\sqrt{\pi } \Phi_{\alpha\beta}^{s -})  - i \sin(\sqrt{\pi } \Phi_{\alpha\beta}^{\rho -}) \cos(\sqrt{\pi } \Theta_{\alpha\beta}^{s +}) \sin(\sqrt{\pi } \Phi_{\alpha\beta}^{s -}) \right ) \\
O_{\rm SDW_z}^{[\alpha\beta]} &=\frac{1}{2}(a_{{\alpha},\uparrow}^{R, \dagger}a_{{\beta},\uparrow}^{L} - a_{{\alpha},\downarrow}^{R, \dagger}a_{{\beta},\downarrow}^{L} ) e^{-i(k_{\alpha}+k_{\beta}) x} - {\alpha} \leftrightarrow {\beta} \notag \\
&\sim  e^{- i (k_{\alpha} + k_{\beta})x} \frac{e^{- i \sqrt{\pi} \Theta_{\alpha\beta}^{\rho +}}}{\pi a}  \left (\cos(\sqrt{\pi } \Phi_{\alpha\beta}^{\rho -}) \cos(\sqrt{\pi } \Theta_{\alpha\beta}^{s +}) \cos(\sqrt{\pi } \Phi_{\alpha\beta}^{s -})  - i \sin(\sqrt{\pi } \Phi_{\alpha\beta}^{\rho -}) \sin(\sqrt{\pi } \Theta_{\alpha\beta}^{s +}) \sin(\sqrt{\pi } \Phi_{\alpha\beta}^{s -}) \right ).
\end{align}
\end{widetext}
\label{eq:Operators}
\end{subequations}

\subsubsection{Interactions}
For the purposes of classifying the instabilities presented in Figs.~\ref{fig:U08J01_RG}-\ref{fig:U08J06_RG}, it is sufficient to keep only those interactions which generate potential terms (e.g. cosine terms) of bosonic fields
\begin{eqnarray}
\mathcal H_{\rm int} &=& -\sum_{\alpha\beta} \frac{\tilde{f}_{\alpha\beta}^\sigma}{2} \sum_{\sigma} a_{{\alpha}, \sigma}^{R,\dagger}a_{{\beta}, \bar \sigma}^{L,\dagger} a_{{\beta}, \sigma}^{L}  a_{{\alpha}, \bar \sigma}^{R}\notag \\
&&+ \sum_{{\alpha \neq \beta}} \sum_{\sigma, \sigma'} \left (\tilde{c}_{\alpha\beta}^\rho + \frac{\tilde{c}_{\alpha\beta}^\sigma}{4}\right )a_{{\alpha}, \sigma}^{R,\dagger}a_{{\alpha, } \sigma'}^{L,\dagger} a_{{\beta}, \sigma'}^{L} a_{{\beta}, \sigma}^{R} \notag \\
&&- \sum_{({\alpha},\sigma) \neq ({\beta},\sigma')} \frac{\tilde{c}_{\alpha\beta}^\sigma}{2} a_{{\alpha}, \sigma}^{R,\dagger}a_{{\alpha, } \sigma'}^{L,\dagger} a_{{\beta}, \sigma}^{L} a_{{\beta}, \sigma'}^{R}, \label{eq:HIA}
\end{eqnarray}
where from now on $\tilde{c}^\sigma_{\alpha\beta}$ etc. are to be understood as the renormalized coupling constants. The notation $\bar \sigma$ means $\downarrow (\uparrow)$ for $ \sigma = \uparrow ( \sigma = \downarrow)$. A bosonization of these terms is presented for each phase separately.

In addition, interactions can generate gradient terms of bosons
\begin{subequations}
\begin{align}
\mathcal H_{\nabla^2} &= \frac{(v_{\alpha} + v_{\beta})\mathcal C_{\alpha\beta}^\rho}{2} \left [ \nabla \Theta_{\alpha}^\rho  \nabla \Theta_{\beta}^\rho - \nabla \Phi_{\alpha}^\rho \nabla \Phi_{\beta}^\rho\right ] \notag \\
&+ \frac{(v_{\alpha} + v_{\beta})\mathcal C_{\alpha\beta}^\sigma}{8} \left [ \nabla \Phi_{\alpha}^s  \nabla \Phi_{\beta}^s-\nabla \Theta_{\alpha}^s \nabla \Theta_{\beta}^s  \right  ] 
\end{align}
with 
\begin{equation}
\mathcal C_{\alpha\beta}^{{\rho, \sigma}} = \left (\begin{array}{ccc}
 c_{\rm I,I}^{{\rho, \sigma}} &  f_{\rm I,II}^{{\rho, \sigma}} &  f_{\rm I,III}^{{\rho, \sigma}} \\ 
 f_{\rm I,II}^{{\rho, \sigma}} &  c_{\rm II,II}^{{\rho, \sigma}} &  f_{\rm II,III}^{{\rho, \sigma}} \\ 
 f_{\rm I,III}^{{\rho, \sigma}} &  f_{\rm II,III}^{{\rho, \sigma}} &  c_{\rm III,III}^{{\rho, \sigma}}
\end{array} \right )_{ \alpha\beta}.
\end{equation}
\end{subequations}

For the RG procedure it is useful to express the contractions of fast fields in terms of a Luttinger parameter matrix
\begin{eqnarray}
\left \langle \Theta_{\alpha}^{{\rho, s}} \Theta_{\beta} ^{{\rho, s}}\right \rangle_{\rm fast }&=& \frac{1}{2\pi} \ln(L/a^*)\underline K^{{\rho, \sigma}}_{\alpha\beta} \\
\left \langle \Phi_{\alpha}^{{\rho, s}} \Phi_{\beta}^{{\rho, s}} \right \rangle_{\rm fast} &=& \frac{1}{2\pi} \ln(L/a^*)[(\underline K^{{\rho,\sigma}})^{-1}]_{\alpha\beta} 
\end{eqnarray}
where $L$ is running length scale. 
\begin{subequations}
\begin{align}
\underline K^{\rho}_{\alpha\beta} &\simeq \delta_{\alpha\beta} - 2\mathcal C_{\alpha\beta}^{\rho}, \\
\underline K^{\sigma}_{\alpha\beta} &\simeq \delta_{\alpha\beta} +\mathcal C_{\alpha\beta}^{\sigma}/2.
\end{align}
\label{eq:Luttingerparameters}
\end{subequations}

\subsubsection{Phases under consideration}

We now investigate each of the four phases obtained from the RG analysis.

\paragraph{C3S2 phase.} The potential part of the bosonized Hamiltonian in the C3S2 phase is
\begin{equation}
\mathcal H_{\rm int} \sim \frac{1}{(2 \pi a_*)^2} c_{\rm I,I}^\sigma \cos(\sqrt{8 \pi} \Theta_{\rm I,s}).
\end{equation}
Since $c_{\rm I,I}^\sigma < 0$, the system locks into one of the minima $\sqrt{2/ \pi} \Theta_{\rm I}^{s} \in \mathbb{Z}$ and thus only two out of three spin modes remain gapless. Comparing with Eqs.~\eqref{eq:Operators}, we readily see that $O_{CDW}^{\rm (I)}$ and $O_{SS}^{\rm (I)}$ have algebraic correlations, while $SDW$ and $TS$ correlations are massive. Fermions near $k_{\rm II, III}$ remain unaffected of the condensation of $\Theta_{\rm I}^s$. 

\paragraph{C2S1a phase.} Using Eqs.~\eqref{eq:CouplingConstantsLightGreen}, Eq.~\eqref{eq:HIA} becomes
\begin{eqnarray}
\mathcal H_{\rm int} &=& -\frac{1}{2}\sum_{{\alpha \neq \beta}} \Big \lbrace \sum_\sigma \tilde{f}_{\alpha\beta}^\sigma a_{{\alpha}, \sigma}^{R,\dagger}a_{{\beta}, \bar \sigma}^{L,\dagger} a_{{\beta}, \sigma}^{L}  a_{{\alpha}, \bar \sigma}^{R} \notag \\
&& + \sum_{\sigma, \sigma'} \tilde c_{\alpha\beta}^\sigma \; a_{{\alpha}, \sigma}^{R,\dagger}a_{{\alpha, } \sigma'}^{L,\dagger} a_{{\beta}, \sigma}^{L} a_{{\beta}, \sigma'}^{R} \Big \rbrace \notag \\
&\sim & - \frac{1}{(2\pi a_*)^2} \sum_{{\alpha \neq \beta}}\Big \lbrace \tilde{f}_{\alpha\beta}^\sigma \; \cos[2 \sqrt{\pi}(\Phi_{\alpha\beta}^{s-} +\Theta_{\alpha\beta}^{s+}) ]\notag \\
&&+\tilde{c}_{\alpha\beta}^\sigma \;  \cos(2 \sqrt{\pi}\Phi_{\alpha\beta}^{\rho -})\cos(2 \sqrt{\pi} \Phi_{\alpha\beta}^{s-}) \notag \\
&&-\tilde{c}_{\alpha\beta}^\sigma \;  \cos(2 \sqrt{\pi}\Phi_{\alpha\beta}^{\rho -})\cos(2 \sqrt{\pi} \Theta_{\alpha\beta}^{s+}) \Big \rbrace. \label{eq:Int1}
\end{eqnarray}
Here we have introduced $\Phi_{\alpha\beta}^{\rho \pm} =
(\Phi_{{\alpha}}^{\rho} \pm\Phi_{{\beta}}^{\rho})/\sqrt{2}$ and
analogous notations for all other channels. Since $\tilde
f_{\alpha\beta}^\sigma \rightarrow - \infty$ and $\tilde
c_{\alpha\beta}^\sigma \rightarrow + \infty$, there are two sets of
solutions which minimize the potential energy for ${\alpha, \beta} \in
\lbrace \text{II,III} \rbrace; {\alpha \neq \beta}$
\begin{subequations}
\begin{align}
& \Phi_{\alpha\beta}^{\rho-}/\sqrt{\pi} \in \mathbb Z , \; \Phi_{\alpha\beta}^{s-}/\sqrt{\pi} \in \mathbb Z, \;\Theta_{\alpha\beta}^{s+}/\sqrt{\pi} \in \mathbb Z +1/2,  \\
& \Phi_{\alpha\beta}^{\rho-}/\sqrt{\pi} \in \mathbb Z +1/2, \;\Phi_{\alpha\beta}^{s-}/\sqrt{\pi} \in \mathbb Z +1/2, \; \Theta_{\alpha\beta}^{s+}/\sqrt{\pi} \in \mathbb Z . 
\end{align}
\label{eq:lightgreenCDW}
\end{subequations}
The low energy theory perturbing about any of the given minima is the same for either solution. Fermions with momenta close to $k_{\rm I}$ remain unaffected.
Only two charge and one spin mode remain gapless, hence the notation C2S1. Comparison with Eqs.~\eqref{eq:Operators} demonstrates that for any of the two solutions of \eqref{eq:lightgreenCDW} the operator $O_{CDW}^{\text{[II,III]}}$ orders. Note that, in view of the locking of $\Phi_{\alpha\beta}^{\rho-}$ into a minimum, the conjugate variable $\Theta_{\alpha\beta}^{\rho-}$ is maximally uncertain and thus $O_{SS}^{\text{[II,III]}}$ and $O_{TS_z}^{\text{[II,III]}}$ do not display long range correlations in either case.

It is instructive to refermionize the interaction term of excitations near Fermi points $\rm II, III$ in the basis of fermions describing fluctuations in the relative charge, relative spin and total spin sectors. Assuming $v_{\rm II} = v_{\rm III}$ for simplicity, Eq.~\eqref{eq:Int1} may be written as
\begin{equation}
 \mathcal H_{\rm int} = - \vert\tilde f^{\sigma}_{\rm II, III}\vert [M_{s^-}M_{s^+} - M_{\rho^-}M_{s^-} + M_{s^+}M_{\rho^-}], \label{eq:GNC2S1a}
\end{equation}
where we introduce mass terms
\begin{equation}
M_a = a^{R,\dagger}_{a} a^{L}_{a} +a^{L,\dagger}_{a} a^{R}_{a}, \quad \text{with } {a} = \rho^-, s^-, s^+.
\end{equation}
If we further perform a gauge transformation in the $(s,+)$ sector, $a^{L}_{s,+} \rightarrow - a^{L}_{s,+}$, Eq.~\eqref{eq:GNC2S1a} corresponds to the interaction term of an SO(6) $\sim$ SU(4) Gross-Neveu model.

\paragraph{C2S1b and C1S0 phases.} Again, we keep only the dominant coupling constants and exploit $c_{\alpha\beta}^\rho = c_{\alpha\beta}^\sigma/4$ for ${\alpha \neq \beta}$. Then 
\begin{align}
\mathcal H_{\rm int} &= - \sum_{{\alpha},\sigma} \frac{\tilde{c}_{{\alpha \alpha}}^\sigma}{2}a_{{\alpha}, \sigma}^{R,\dagger}a_{{\alpha}, \bar \sigma}^{L,\dagger} a_{{\alpha}, \sigma}^{L}  a_{{\alpha}, \bar \sigma}^{R}\notag \\
&-\sum_{{\alpha \neq \beta}} \sum_{\sigma}  \frac{\tilde{c}_{\alpha\beta}^\sigma}{2}  a_{{\alpha}, \sigma}^{R,\dagger}a_{{\alpha, } \bar \sigma}^{L,\dagger} (a_{{\beta}, \sigma}^{L} a_{{\beta}, \bar \sigma}^{R}- a_{{\beta}, \bar \sigma}^{L} a_{{\beta},  \sigma}^{R})  \notag \\
& \sim \frac{1}{(2\pi a_*)^2} \Big \lbrace \tilde{c}_{{\alpha \alpha}}^\sigma \cos(\sqrt{8 \pi} \Theta_{{\alpha}}^s) \notag \\
&+ 4 \sum_{{\alpha < \beta}} \tilde c_{\alpha\beta}^\sigma \cos(\sqrt{4\pi} \Phi_{\alpha\beta}^{\rho-}) \cos(\sqrt{2\pi} \Theta_{\alpha}^s) \cos(\sqrt{2\pi} \Theta_{\beta}^s) \Big \rbrace. \label{eq:BosonC2S1b}
\end{align}
We note that in the C2S1b phase, $\tilde{c}_{\alpha\beta}^\sigma \rightarrow - \infty$ for ${\alpha, \beta} = \text{II,III}$. Thus the minimum of the potential is 
\begin{equation}
\sqrt{2/ \pi} \Theta_{\rm II,s} \in \mathbb{Z}, \; \sqrt{2/ \pi} \Theta_{\rm III,s} \in \mathbb{Z}, \; \Phi_{\rm II, III}^{\rho -}/\sqrt{\pi} \in \mathbb{Z}.
\end{equation}
Thus the C2S1b phase has two gapless charge modes and one gapless spin
mode. It is, in essence, a spinful Luttinger liquid near Fermi point $k_{\rm I}$ and
a superconductor with equal gaps ($\tilde c_{\rm II,II}^\sigma = \tilde
c_{\rm III,III}^\sigma$) at Fermi points $k_{\rm II,III}$. Again, we
can refermionize the interaction term of the C2S1b phase in the same
channels as in the case of C2S1a. At $v_{\rm II} = v_{\rm
III}$ we obtain

\begin{equation}
\mathcal H_{\rm int} = -2 |\tilde c_{\rm II, II}|[M_{s^+}M_{s^-} + M_{\rho^-}(M_{s^+} + M_{s^-})]
\end{equation}
which represents an SO(6) Gross-Neveu model, albeit in a different phase then in the case of C2S1a.

The C1S0 phase is characterized by $\tilde{c}_{{\alpha \alpha}}^\sigma \rightarrow - \infty$, $\tilde{c}_{\rm II, III}^\sigma \rightarrow - \infty$ and $\tilde{c}_{\rm I, II}^\sigma \rightarrow  \infty$, $\tilde{c}_{\rm I,III}^\sigma \rightarrow \infty$. Therefore, the minimum occurs at
\begin{align}
&\sqrt{2/ \pi} \Theta_{{\alpha},s} \in \mathbb{Z}, \; \Phi_{\rm I, II}^{\rho -}/\sqrt{\pi} \in \mathbb{Z} + 1/2, \\
&\Phi_{\rm I, III}^{\rho -}/\sqrt{\pi} \in \mathbb{Z} + 1/2,\; \Phi_{\rm II, III}^{\rho -}/\sqrt{\pi} \in \mathbb{Z}.
\end{align}
Since $\Phi_{\rm I, II}^{\rho -} = \Phi_{\rm I, III}^{\rho -} -
\Phi_{\rm II, III}^{\rho -}$ there are two independent constraints on
bosons in the charge sector and three independent constraints on
bosons in the spin sector, justifying the notation C1S0. This phase is a fully
gapped spin singlet $s_{+--}$ intraband superconductor with the
following products of gap functions: $\Delta_{\rm I} \Delta_{\rm
II}<0, \Delta_{\rm I} \Delta_{\rm III}<0,\Delta_{\rm II} \Delta_{\rm
III}>0$.

\subsection{Unequal interorbital and intra-orbital repulsion}
\label{app:NonEqualHubbard}

This appendix examines the effect of unequal inter and intra-orbital repulsion, i.e. $\tilde J \neq 0$ in
Eqs.~\eqref{eq:bareinteraction},\eqref{eq:bareinteractioncorrections}. We
note that at $\tilde J = J$,
the Hubbard-Kanamori interaction 
takes
the form
\begin{align}
H_{U} ({j})&= \frac{U}{2}  { \sum_{\substack{\tau,  {\gamma, \sigma,}\\ { \gamma', \sigma'}}}}' n_{ \tau \gamma\sigma} ({j}) n_{\tau \gamma' \sigma'} ({j}) \notag \\
&+ 2 J \sum_{\tau} \lbrace  [T^{(x)}_{\tau}(j)]^2+[T^{(y)}_{\tau}(j) ]^2 \rbrace.
\end{align}
Here, $T^{(\mu)}_\tau(j) = d^\dagger_{\sigma, \tau} \tilde \tau_\mu d_{\sigma, \tau}/2$ is the orbital isosopin operator and $\tilde \tau_\mu$ are Pauli matrices in orbital space.

The integration of RG equations for general $0\leq J/U \leq 0.7$ and
$0 \leq \tilde J/J < 1.2$ reveals five phases and a rather extended
critical regime (see Fig.~\ref{fig:GeneralHund}). In addition to the
four phases discussed in the main text there is an extended
critical regime corresponding to the C3S3 QCP of
Fig.~\ref{fig:SchematicRG} where the numerical integration of RG
(consistently performed at $U = 5 \bar v$) does not reveal  a
divergence for any $y < 1000$ (see Fig.~\ref{fig:Critical}). This
corroborates the finding summarized in Fig.~\ref{fig:SchematicRG} and
highlights the importance of the critical phase.

A typical RG flow for the phase C2S1c is shown in
Fig.~\ref{fig:C2S1cRG}. Among the data points of
Fig.~\ref{fig:GeneralHund} which fall into the C2S1c phase, $\tilde J
= 1.2 J, J = 0.5 U$ has the highest $T_c \sim 0.1 mK$ for $\Lambda = 1
eV$ at $U/\bar v = 5$. This phase is characterized by $c_{\rm
I,I}^\sigma \simeq c_{\rm I,I}^\sigma\simeq - 4 c_{\rm I, II}^\rho
\simeq -c_{\rm I,II}^\sigma \rightarrow -\infty$ with $c_{\rm
I,I}^\rho \simeq c_{\rm II,II}^\rho = - f_{\rm I,II}^\rho \rightarrow
-\infty$. We now can exploit Eq.~\eqref{eq:BosonC2S1b} for $\alpha,
\beta \in \lbrace \text{I,II} \rbrace$, revealing that the minimum
given by
\begin{equation}
\sqrt{2/ \pi} \Theta_{\rm I,s} \in \mathbb{Z}, \; \sqrt{2/ \pi} \Theta_{\rm II,s} \in \mathbb{Z}, \; \Phi_{\rm I, II}^{\rho -}/\sqrt{\pi} \in \mathbb{Z} + 1/2.
\end{equation}
describes a two band superconductor with relative phase $\pi$. 

\begin{figure}
\includegraphics[scale=.4]{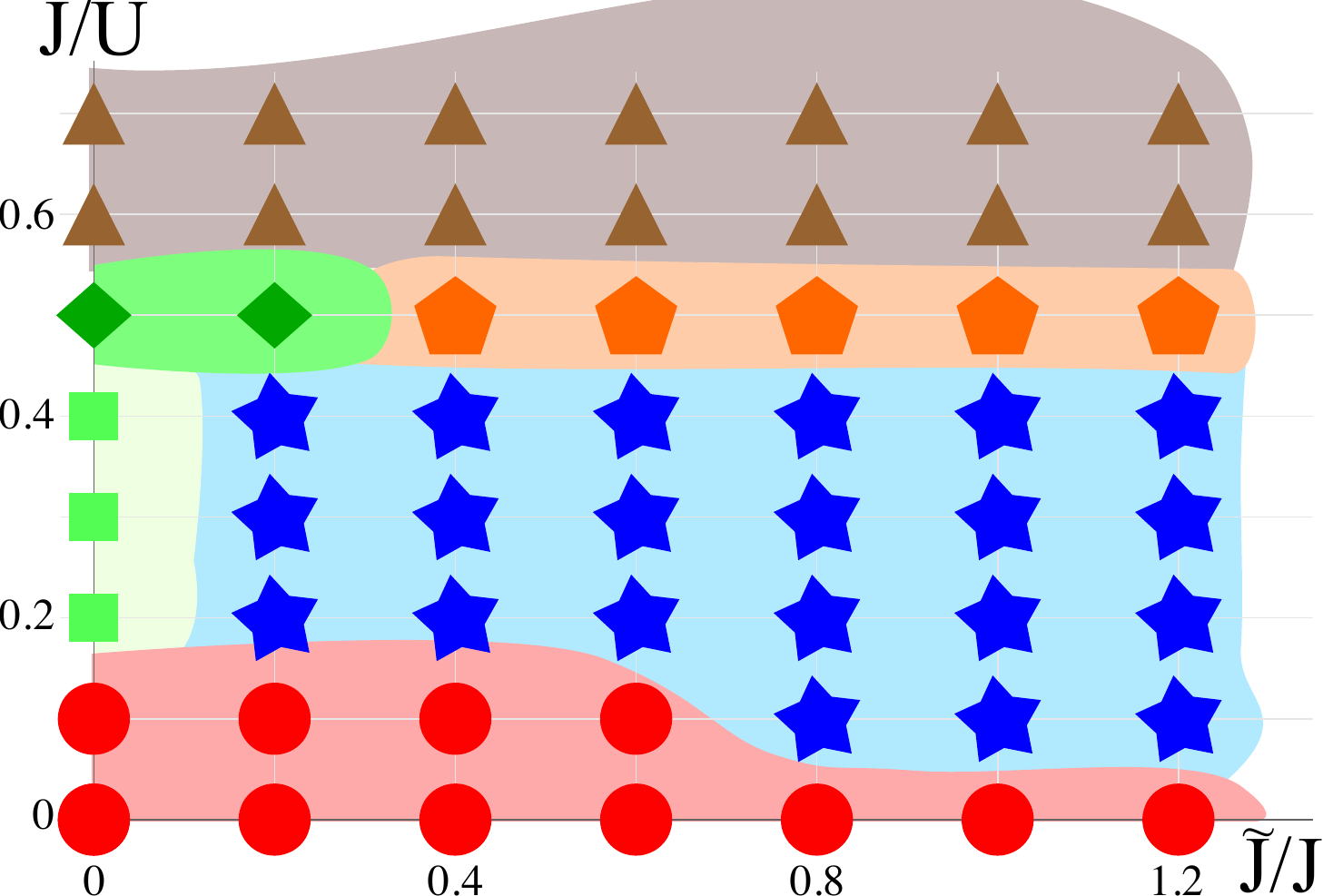} 
\caption{Summary of the integration of RG equations at finite $\tilde
J/J$ and $U/\bar v = 5$. In addition to the phases C3S2 (red dots),
C2S1a (light green squares), C2S1b (dark green diamonds), C1S0 (brown
triangles) we find an additional phase C2S1c (orange pentagons). For
all of those phases the divergence occurs at running scales $y_c <
100$ (for C1S0 and C3S2 $y_c <27$) while in the extended critical region (blue stars), no divergence occurs for any $y \leq 1000$.}
\label{fig:GeneralHund}
\end{figure}

\begin{figure}
\includegraphics[scale=.4]{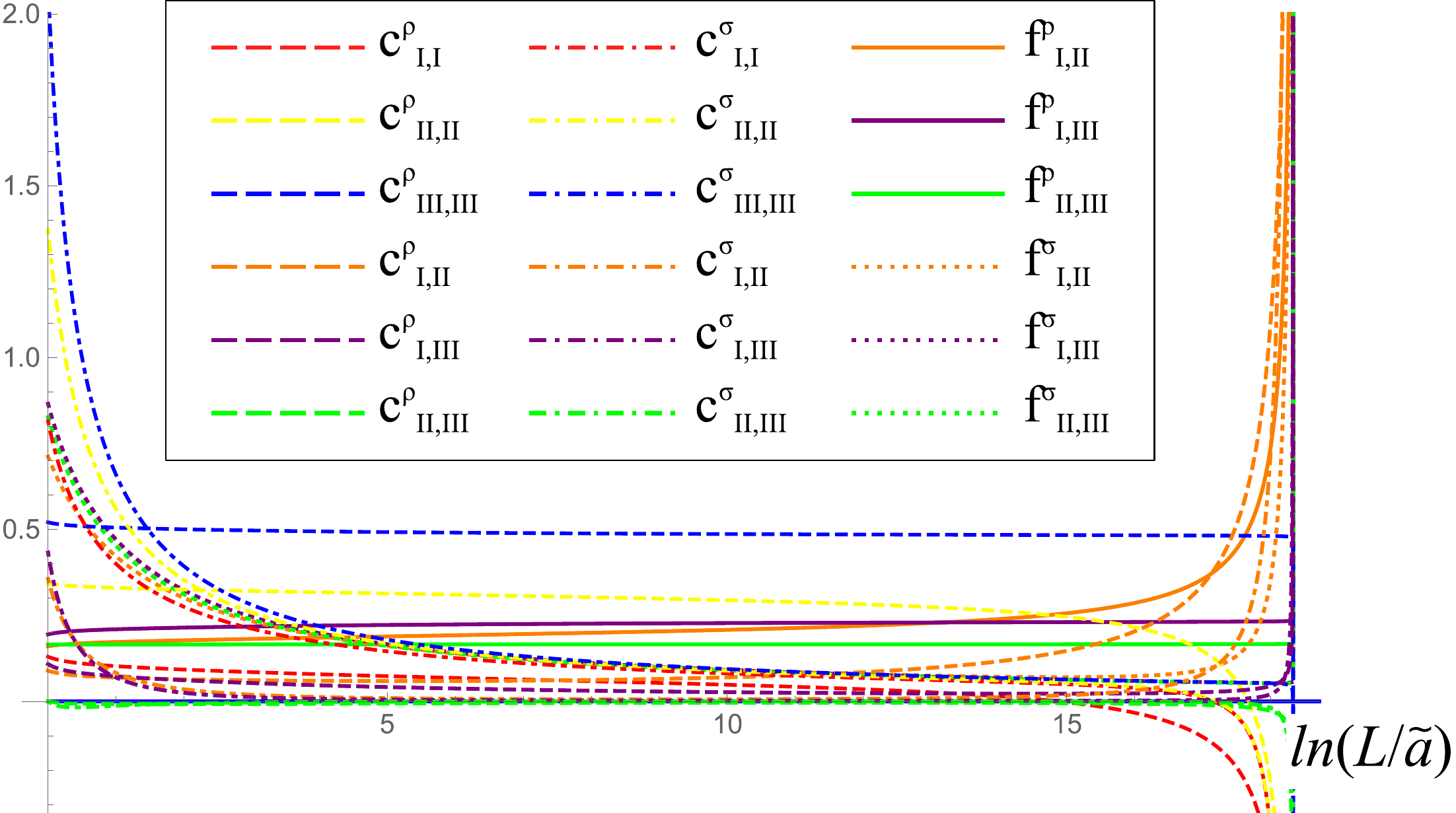} 
\caption{Numerical integration of RG equations for starting values determined by $U/\bar v = 5$ and $J/U = 0.5$, $\tilde J = 1.2 J$. In this case, the system flows to the C2S1c phase.}
\label{fig:C2S1cRG}
\end{figure}

\begin{figure}
\includegraphics[scale=.4]{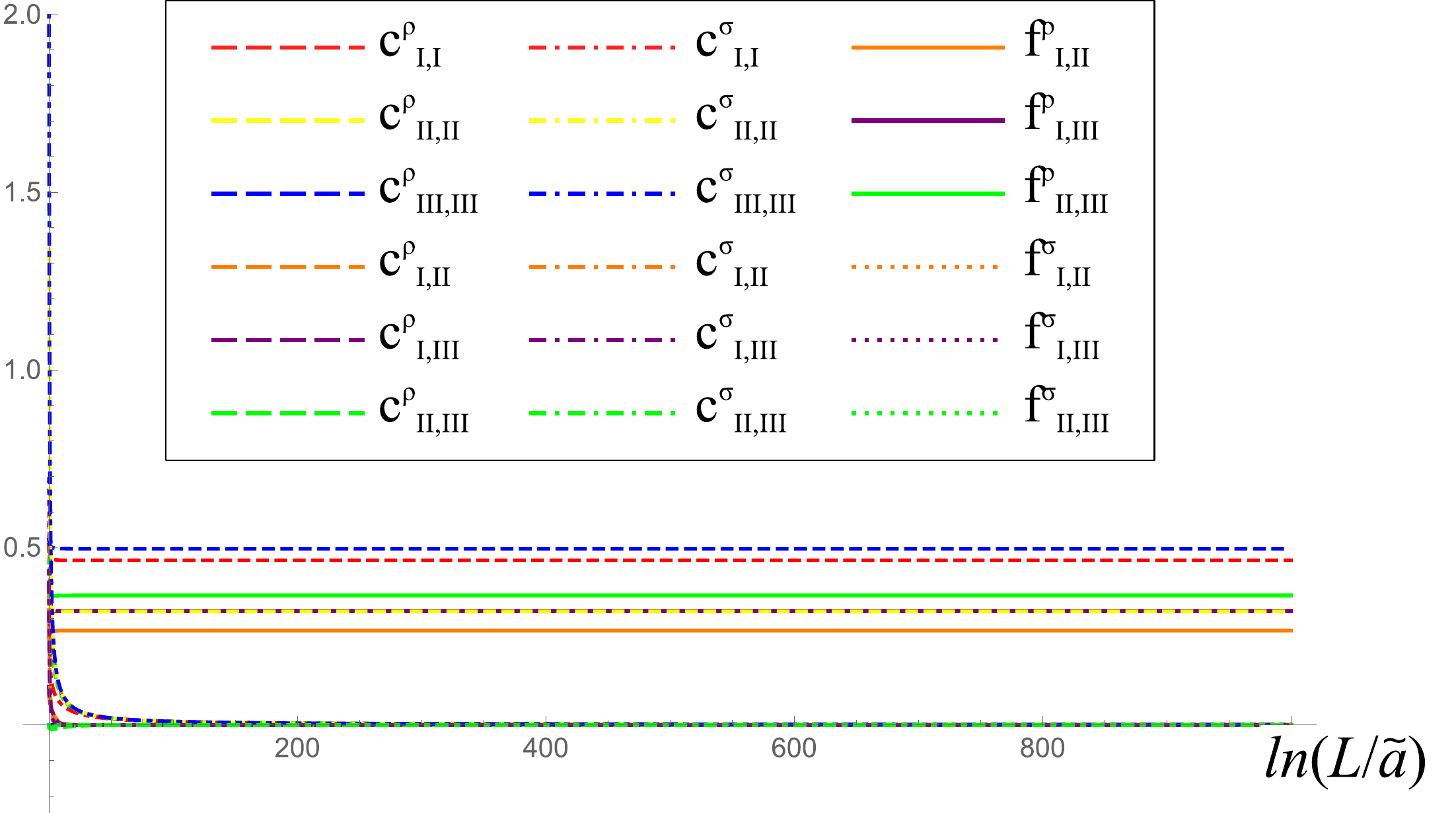} 
\caption{Numerical integration of RG equations for starting values determined by $U/\bar v = 5$ and $J/U = 0.3$ and $\tilde J = 0.4 J$. In this case, the system remains critical for any $y \leq 1000$.}
\label{fig:Critical}
\end{figure}

\section{Umklapp scattering}
\label{app:umklapp}

In this appendix we provide details on umklapp scattering as a discussed in Sec.~\ref{sec:Umklapp}. 

\subsection{Bosoniziation of umklapp terms}
We consider
\begin{equation}
\mathcal H_{\rm u} = - G_u (2\pi a) \Big [ \left (\prod_{{\alpha} =\rm I}^{\rm III} a_{{\alpha, }\sigma'_{\alpha}}^{R, \dagger} a_{{\alpha, }\sigma_{\alpha}}^L\right ) \delta_{\lbrace \sigma' \rbrace, \lbrace\sigma \rbrace} \Big ] + H.c.. \label{eq:umklappAppendix}
\end{equation}
The symbol $\delta_{\lbrace \sigma'\rbrace, \lbrace\sigma \rbrace}$
implies equality of the two sets $\lbrace \sigma'_{\rm I}\sigma'_{\rm
II}\sigma'_{\rm III} \rbrace$ and $\lbrace \sigma_{\rm I}\sigma_{\rm
II}\sigma_{\rm III} \rbrace$ of spin indices, which reflects the
overall spin conservation. Note that the spin is not conserved within
any  given pair of Fermi points, i.e. in general $\sigma_{\alpha} \neq
\sigma_{\alpha}'$ (see Fig.~\ref{fig:Umklapp}). A summation over all
permutations of spin indices which preserve the overall spin
conservation is implied. For this section we therefore concentrate on
the term which is fully symmetric under exchange of spin indices, more
generic terms are discussed afterwards.

Bosonization of Eq.~\eqref{eq:umklappAppendix} leads to 
\begin{eqnarray}
\mathcal H_{\rm u} &=& g_u e^{-i \sqrt{2\pi} \sum_{{\alpha} =\rm I}^{\rm III}\Theta^\rho_{{\alpha}}} \notag \\
&\times &\prod_{{\alpha} =\rm I}^{\rm III} \left [\cos  (\sqrt{2\pi} \Theta_{\alpha}^s) - i \sin  (\sqrt{2\pi} \Phi_{\alpha}^s)\right ] + H.c..
\end{eqnarray}
In principle, $g_u = i 2G_u/(\pi a)^2$ can have both real and imaginary parts. This leads to an overall of 16 umklapp terms, all of which may have different bare values in the case of spin dependent $G_u$. In addition, under RG, interband interaction generates additional terms.

\subsection{Analysis of umklapp scattering}

We first analyze umklapp scattering in the Luttinger phase prior to
an instability, proveeding to  each of the phases
obtained above. Employing fermionic diagrams, Fig.~\ref{fig:Umklapp},
illustrates that the three-body umklapp scattering can not renormalize
two-body interactions at weak coupling in the one-loop
approximation. Therefore we can determine the scaling dimension $d_u$
of the most dominant umklapp process without considering its
backreaction on the other coupling constants.

\paragraph{Umklapp scattering in the Luttinger liquid phase}

When $\mathcal C_{\alpha\beta}^\sigma$ has predominantly positive
entries and $\underline K^{\sigma}$ has eigenvalues large than unity,
the umklapp terms with the largest dimension involve $\Phi^s_{\alpha}$,
so we disregard all terms with $\Theta^s_{\alpha}$. We introduce
$\Theta_{\rm tot}^\rho = [\sum_{\alpha}
\Theta_{\alpha}^\rho]/\sqrt{3}$
\begin{align}
\mathcal H_{\rm u} &= \cos(\sqrt{6\pi}\Theta_{\rm tot}^\rho) \times \notag \\
&\times \Big [ g_{sss} \sin(\sqrt{2\pi} \Phi_{\rm I}^s)\sin(\sqrt{2\pi} \Phi_{\rm II}^s)\sin(\sqrt{2\pi} \Phi_{\rm III}^s) \notag \\
&+g_{scc} \sin(\sqrt{2\pi }\Phi_{\rm I}^s)\cos(\sqrt{2\pi }\Phi_{\rm II}^s)\cos(\sqrt{2\pi }\Phi_{\rm III}^s) \notag \\
&+g_{csc} \cos(\sqrt{2\pi} \Phi_{\rm I}^s)\sin(\sqrt{2\pi} \Phi_{\rm II}^s)\cos(\sqrt{2\pi }\Phi_{\rm III}^s) \notag \\
&+g_{ccs} \cos(\sqrt{2\pi} \Phi_{\rm I}^s)\cos(\sqrt{2\pi} \Phi_{\rm II}^s)\sin(\sqrt{2\pi }\Phi_{\rm III}^s) \Big ].
\end{align}
By contrast, when $\mathcal C_{\alpha\beta}^\sigma$ has
(predominantly) negative eigenvalues we disregard terms with
$\Phi^s_{\alpha}$ and keep
\begin{align}
\mathcal H_{\tilde{\rm u}} &= \cos(\sqrt{6\pi}\Theta_{\rm tot}^\rho) \times \notag \\
&\times \Big [\tilde g_{ccc} \cos(\sqrt{2\pi} \Theta_{\rm I}^s)\cos(\sqrt{2\pi} \Theta_{\rm II}^s)\cos(\sqrt{2\pi} \Theta_{\rm III}^s) \notag \\
&+\tilde g_{css} \cos(\sqrt{2\pi} \Theta_{\rm I}^s)\sin(\sqrt{2\pi} \Theta_{\rm II}^s)\sin(\sqrt{2\pi} \Theta_{\rm III}^s) \notag \\
&+\tilde g_{scs} \sin(\sqrt{2\pi} \Theta_{\rm I}^s)\cos(\sqrt{2\pi} \Theta_{\rm II}^s)\sin(\sqrt{2\pi} \Theta_{\rm III}^s) \notag \\
&+\tilde g_{ssc} \sin(\sqrt{2\pi} \Theta_{\rm I}^s)\sin(\sqrt{2\pi} \Theta_{\rm II}^s)\cos(\sqrt{2\pi} \Theta_{\rm III}^s) \Big ].
\end{align}
In both cases, terms with $\cos(\sqrt{6\pi}\Theta_{\rm tot}^\rho)
\rightarrow \sin(\sqrt{6\pi}\Theta_{\rm tot}^\rho) $ may also
exist. They have the same scaling dimension as the cosine terms shown here.

The tree level RG equations are
\begin{widetext}
\begin{align}
\left (\begin{array}{c}
g_{sss} \\ 
g_{ssc} \\ 
g_{csc} \\ 
g_{ccs}
\end{array} \right )^{\mathbf{.}} &= \left  [2 - \frac{3 K_{\rm tot}^\rho + \sum_{\alpha} (\underline K^{\sigma,-1})_{{\alpha \alpha}}}{2} + \left (\begin{array}{cccc}
0 & (\underline{K}^{\sigma,-1})_{\rm II,III} & (\underline{K}^{\sigma,-1})_{\rm I,III} & ( \underline{K}^{\sigma,-1})_{\rm I,II} \\  
(\underline{K}^{\sigma,-1})_{\rm II,III} & 0 & -(\underline{K}^{\sigma,-1})_{\rm I,II} & -(\underline{K}^{\sigma,-1})_{\rm I,III} \\  
( \underline{K}^{\sigma,-1})_{\rm I,III} & -(\underline{K}^{\sigma,-1})_{\rm I,II} & 0 & -(\underline{K}^{\sigma,-1})_{\rm II,III} \\ 
( \underline{K}^{\sigma,-1})_{\rm I,II} & -(\underline{K}^{\sigma,-1})_{\rm I,III} & -(\underline{K}^{\sigma,-1})_{\rm II,III} & 0
\end{array} \right )\right  ]\left (\begin{array}{c}
g_{sss} \\ 
g_{ssc} \\ 
g_{csc} \\ 
g_{ccs}
\end{array} \right ) \label{eq:umklappRGPhi} \\
\left (\begin{array}{c}
\tilde g_{ccc} \\ 
\tilde g_{ccs} \\ 
\tilde g_{scs} \\ 
\tilde g_{ssc}
\end{array} \right )^{\mathbf{.}} &= \left  [2 - \frac{3 K_{\rm tot}^\rho + \sum_{\alpha} (\underline K^{\sigma})_{{\alpha \alpha}}}{2} + \left (\begin{array}{cccc}
0 & (\underline{K}^{\sigma})_{\rm II,III} & (\underline{K}^{\sigma})_{\rm I,III} & ( \underline{K}^{\sigma})_{\rm I,II} \\  
(\underline{K}^{\sigma})_{\rm II,III} & 0 & -(\underline{K}^{\sigma})_{\rm I,II} & -(\underline{K}^{\sigma})_{\rm I,III} \\  
( \underline{K}^{\sigma})_{\rm I,III} & -(\underline{K}^{\sigma})_{\rm I,II} & 0 & -(\underline{K}^{\sigma})_{\rm II,III} \\ 
( \underline{K}^{\sigma})_{\rm I,II} & -(\underline{K}^{\sigma})_{\rm I,III} & -(\underline{K}^{\sigma})_{\rm II,III} & 0
\end{array} \right )\right  ]\left (\begin{array}{c}
\tilde g_{ccc} \\ 
\tilde g_{ccs} \\ 
\tilde g_{scs} \\ 
\tilde g_{ssc}
\end{array} \right )\label{eq:umklappRGTheta}
\end{align}
\end{widetext}

At the bare level interections in the spin sector are weak and
repulsive $\mathcal C_{\alpha\beta}^\sigma >0$, renormalize downwards as
the QCP is approached (see Fig.~\ref{fig:SchematicRG}). The Luttinger
parameter in the total charge sector is
\begin{equation}
K_{\rm tot}^\rho = \frac{\sum_{\alpha\beta} \underline K^\rho_{\alpha\beta}}{3}.
\end{equation}
The largest scaling dimension, which occurs in
Eq.~\eqref{eq:umklappRGPhi}, is typically negative and as $[\underline
K^\sigma]_{\alpha\beta}\rightarrow \delta_{\alpha\beta}$ from above,
becomes $[1-3 K_{\rm tot}^\rho]/2$. Thus, a three band {LL} with weak
interactions does not display Mott localization.

\paragraph{Umklapp scattering in the C3S2 phase.}

In the C3S2 phase, $\Theta_{\rm I}^s$ condenses, while the Luttinger
parameter in the sector of channels II, III remains positive and
approximately $ c^\sigma_{\rm II,II} = c^\sigma_{\rm III,III} =
f^\sigma_{\rm II,III} \ll 1$ reflecting the spin symmetry being
enhanced at the interband level. The dominant umklapp terms are
\begin{align}
\mathcal H_{\rm u} &= \cos(\sqrt{6\pi}\Theta_{\rm tot}^\rho) \times \notag \\
&\times \Big [ g_{cc} \cos(\sqrt{2\pi }\Phi_{\rm II}^s)\cos(\sqrt{2\pi }\Phi_{\rm III}^s) \notag \\
&+ g_{ss} \sin(\sqrt{2\pi} \Phi_{\rm II}^s)\sin(\sqrt{2\pi} \Phi_{\rm III}^s).
\end{align}
In this phase, there are three other pairs of operators which have the same RG equations, namely
\begin{align}
\left (\begin{array}{c}
g_{cc} \\ 
g_{ss}
\end{array} \right )^{\mathbf{.}} &= \Bigg  [2 - \frac{3 K_{\rm tot}^\rho + \sum_{{\alpha, }\rm II}^{\rm III} (\underline K^{\sigma,-1})_{{\alpha \alpha}}}{2} \notag \\
&+ \left (\begin{array}{cc}
0 & (\underline{K}^{\sigma,-1})_{\rm II,III}  \\  
(\underline{K}^{\sigma,-1})_{\rm II,III} & 0 
\end{array} \right )\Bigg  ]\left (\begin{array}{c}
g_{cc} \\ 
g_{ss}
\end{array} \right )
\end{align}
The dominant operator is obtained for $g_{cc} = g_{ss}$ and has
scaling dimension $2 - \frac{3 K_{\rm tot}^\rho + \sum_{{\alpha, }\rm
II}^{\rm III} (\underline K^{\sigma,-1})_{{\alpha
\alpha}}-2(\underline K^{\sigma,-1})_{\rm II,III}}{2} \simeq 1
-\frac{3 K_{\rm tot}^\rho}{2}$. Therefore, the Mott transition occurs
at $K_{\rm tot} = 2/3$ which requires rather stong interactions.

\paragraph{Umklapp scattering and the phase C2S1a.}

In this phase the spin sector of channels II, III is fully gapped and $0< c_{\rm I,I}^\sigma \ll \bar v$. 
We therefore concentrate on an umklapp term for which there is spin conservation within Fermi surface I, i.e.
\begin{equation}
\mathcal H_{\rm u} = g \cos(\sqrt{6\pi}\Theta_{\rm tot}^\rho) \cos(\sqrt{2\pi }\Theta_{\rm I}^s). \label{eq:umklappC2S1a}
\end{equation}
and the analogous term obtained by $\cos(\sqrt{6\pi}\Theta_{\rm tot}^\rho) \rightarrow \sin(\sqrt{6\pi}\Theta_{\rm tot}^\rho)$. The scaling dimention of these terms is $2 - {3 K_{\rm tot}^\rho}/{2} - K_{\rm I,I}^\sigma/2$ and thus the transition occurs at 
\begin{equation}
K_{\rm tot}^\rho = \frac{2}{3} \left [2 - \frac{K_{\rm I,I}^\sigma}{2 }\right ] \approx 1. 
\end{equation}
We observe that the ordering in the spin sector promotes a Mott
transition in its vicinity and we expect $K_{\rm tot}^\rho <1$. 
When the Mott transition occurs, the total charge mode
$\Theta_{\rm tot}^\rho$ freezes,  corresponding to an electrical charge
insulator. At the same time $\Theta_{\rm I}^s$ freezes. Taken
together,  the phase phase C2S1a becomes a phase C1S0a.

The long-range correlations of $\mathcal O^{[\rm II, III]}_{\rm CDW}$ of the C2S1a phase survive the Mott transition and additional long-range correlations of $\mathcal O^{\rm I, I}_{\rm CDW}$ appear. 

\paragraph{Umklapp scattering and the phase C2S1b.}

The umklapp terms of relevance for the phase C2S1b are the also given
by Eq.~\eqref{eq:umklappC2S1a} and the transition to a phase C1S0b
again occurs at $K_{\rm tot}^\rho \approx 1$. All superconducting
correlations are killed by the ordering of $\Theta_{\rm tot}^\rho$,
the only long range correlations occur for $\mathcal O^{\rm I, I}_{\rm
CDW}$.

\paragraph{Umklapp scattering and the phase C1S0.}

\begin{figure}[t]
\includegraphics[scale=.6]{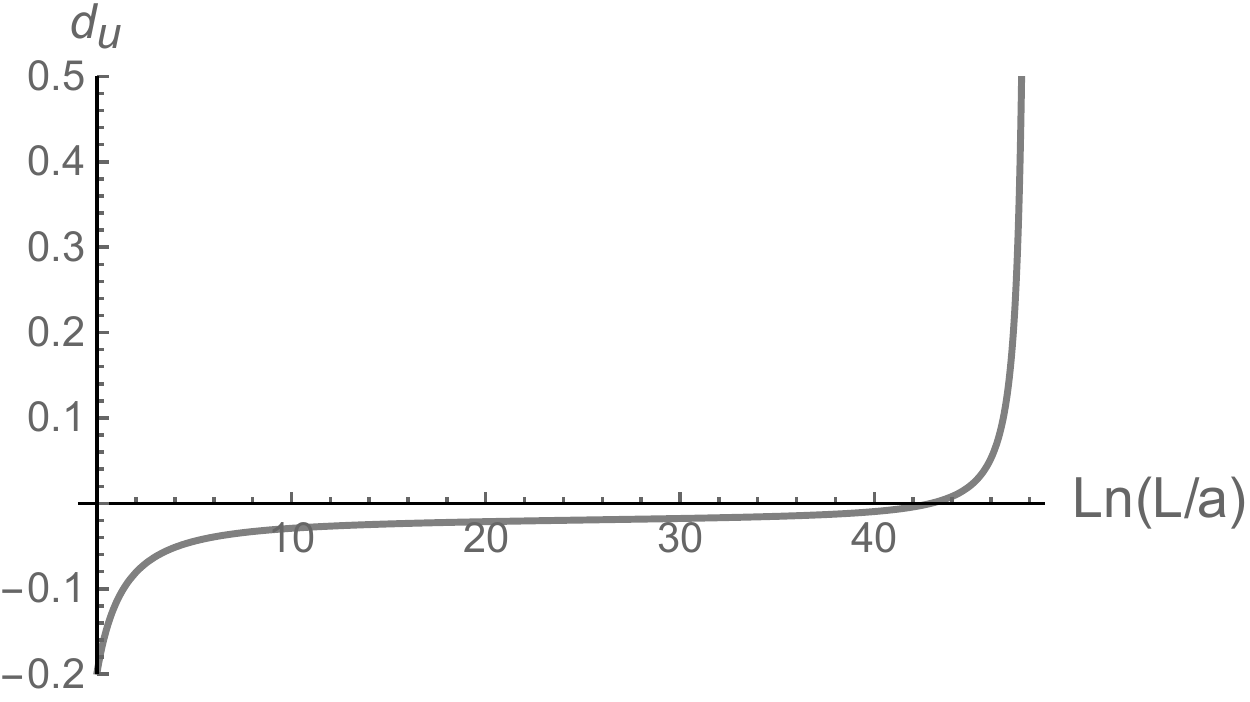} 
\caption{Scaling dimension $d_u$, see Eq.~\eqref{eq:umklappDim}, for the umklapp operator given in Eq.~\eqref{eq:umklappC1S0} as a function of RG-time. Here, $J/U = 0.6$ and $U =1.5\bar v$ were assumed and we used Eqs.~\eqref{eq:Luttingerparameters} for the evaluation of the Luttinger parameters. Clearly, the umklapp operator becomes relevant before the transition to the superconducting state at $\ln(L/\tilde a) = 48$.}
\label{fig:umklappDim}
\end{figure}

In the phase C1S0 all spin modes are frozen and only the pair $\Theta_{\rm tot}^\rho, \Phi_{\rm tot}^\rho$ displays long-range correlations. The umklapp term is 

\begin{equation}
\mathcal H_{\rm u} = g \cos(\sqrt{6\pi}\Theta_{\rm tot}^\rho),
\end{equation}
which has dimension $2 - 3K_{\rm t
ot}^\rho/2$ and thus appears to be relevant for $K_{\rm tot}^\rho <4/3$, i.e. even for attractive interactions. This seems physically inconsistent and a more appropriate treatment of the umklapp scattering for the phase C1S0 follows.

In the present case where $\mathcal C_{{\alpha \alpha}}^\sigma
\rightarrow - \infty$ for all three diagonal matrix elements, the Mott
transition at half filling occurs prior to the instability to the
fully gapped superconductor. Indeed, as we see from
Eq.~\eqref{eq:umklappRGTheta}, the dominant operator has the form
\begin{align}
\mathcal {H}_{{\rm u}} &= g \cos(\sqrt{6\pi}\Theta_{\rm tot}^\rho) \times \notag \\
&\times \Big [ \cos(\sqrt{2\pi} \Theta_{\rm I}^s)\cos(\sqrt{2\pi} \Theta_{\rm II}^s)\cos(\sqrt{2\pi} \Theta_{\rm III}^s) \notag \\
&- \cos(\sqrt{2\pi} \Theta_{\rm I}^s)\sin(\sqrt{2\pi} \Theta_{\rm II}^s)\sin(\sqrt{2\pi} \Theta_{\rm III}^s) \notag \\
&+ \sin(\sqrt{2\pi} \Theta_{\rm I}^s)\cos(\sqrt{2\pi} \Theta_{\rm II}^s)\sin(\sqrt{2\pi} \Theta_{\rm III}^s) \notag \\
&+ \sin(\sqrt{2\pi} \Theta_{\rm I}^s)\sin(\sqrt{2\pi} \Theta_{\rm II}^s)\cos(\sqrt{2\pi} \Theta_{\rm III}^s) \Big ] \notag \\
& = g \cos(\sqrt{6\pi}\Theta_{\rm tot}^\rho)\cos(\sqrt{6\pi}\Theta_{\rm rel}^s) \label{eq:umklappC1S0}
\end{align}
with $\sqrt{3}\Theta_{\rm rel}^s = \Theta_{\rm I}^s - \Theta_{\rm II}^s - \Theta_{\rm III}^s$. The scaling dimension of this operator is
\begin{equation}
d_u= 2 - 3\frac{K_{\rm tot}^\rho + K_{\rm rel}^\sigma}{2}, \label{eq:umklappDim}
\end{equation}
where 
\begin{equation}
K_{\rm rel}^\sigma = \frac{\sum_{{\alpha}}K_{{\alpha \alpha}}^\sigma + 2 K_{\rm II, III}^\sigma - 2 \sum_{\rm {\alpha} = II, III}K_{\rm I, {\alpha}}^\sigma}{3}.
\end{equation}
When the operator Eq.~\eqref{eq:umklappC1S0} orders prior to the C1S0 instability, an insulating C2S2 emerges, see Fig.~\ref{fig:umklappDim}.


%

\end{document}